\documentclass[a4paper,preprint,superscriptaddress,nofootinbib,longbibliography]{revtex4-1}

\usepackage{graphicx}
\usepackage{subfigure} 
\usepackage{amsmath}
\usepackage{geometry}
\usepackage{multirow}
\usepackage{hyperref}
\usepackage{cancel}
\usepackage{color}
\usepackage{diagbox}

\begin{document}

\title{Doubly charged Higgs boson at same-sign lepton colliders}

\author{Cheng-Wei Chiang}
\email{chengwei@phys.ntu.edu.tw}
\affiliation{Department of Physics and Center for Theoretical Physics, National Taiwan University, Taipei
10617, Taiwan}
\affiliation{Physics Division, National Center for Theoretical Sciences, Taipei 10617, Taiwan}

\author{Kazuki Enomoto}
\email{k\_enomoto@phys.ntu.edu.tw}
\affiliation{Department of Physics and Center for Theoretical Physics, National Taiwan University, Taipei
10617, Taiwan}

\author{Min-Yuan Liao}
\email{r13222088@ntu.edu.tw}
\affiliation{Department of Physics and Center for Theoretical Physics, National Taiwan University, Taipei
10617, Taiwan}

\begin{abstract}
We investigate the search for doubly charged Higgs bosons in the Georgi-Machacek (GM) model at same-sign lepton colliders. 
The dominant production mode is vector boson fusion, through which the particle is singly produced and decays into a pair of same-sign $W$ bosons if the triplet vacuum expectation value is sufficiently large, as allowed in the GM model. 
Considering the leptonic decays of the $W$ bosons, we discuss the discovery reach of the signal and a method to extract the mass of the doubly charged Higgs boson. 
It is impossible to construct the transverse mass of the decay product to obtain the mass because of the neutrinos radiated from the initial-state leptons. 
Alternatively, we propose to make use of the invariant mass of the same-sign leptons.  
We perform  $\chi^2$ fitting using the analytical formula for the invariant mass distribution and demonstrate that this approach is effective in extracting the information on the mass.  We also compare the process with pair production processes at the opposite-sign lepton colliders.
\end{abstract}

\maketitle

\section{Introduction}

An isospin triplet scalar field with the hypercharge $Y=1$ is often introduced in physics beyond the standard model (SM), such as the Higgs triplet model (HTM)~\cite{Konetschny:1977bn, Cheng:1980qt, Schechter:1980gr, Magg:1980ut, Lazarides:1980nt, Mohapatra:1980yp}, the Georgi-Machacek (GM) model~\cite{Chanowitz:1985ug, Georgi:1985nv}, and the littlest Higgs model~\cite{Arkani-Hamed:2002ikv}. 
The triplet field plays a crucial role in explaining certain phenomena in particle physics and cosmology, such as tiny neutrino masses~\cite{Konetschny:1977bn, Cheng:1980qt, Schechter:1980gr, Magg:1980ut, Lazarides:1980nt, Mohapatra:1980yp} and baryon asymmetry of the Universe~\cite{Barrie:2021mwi, Barrie:2022cub}.

A distinctive feature of such models is the existence of the doubly charged Higgs boson from the triplet.\footnote{There are other candidates of the doubly charged Higgs boson: a singlet with $Y=2$ or a part of a doublet with $Y=3/2$. Their phenomenology has been investigated in Refs.~\cite{Zee:1985id, Babu:1988ki, Gustafsson:2012vj, Gustafsson:2014vpa, King:2014uha, Sugiyama:2012yw, Nomura:2017abh, Cheng:1980qt, Gunion:1995mq, Rentala:2011mr, Alloul:2013raa, Cheung:2017kxb} (singlet) and Refs.~\cite{Aoki:2011yk, Okada:2015hia, Enomoto:2019mzl, Das:2020pai, Enomoto:2021fql, Cheung:2017kxb, Gunion:1995mq, Rentala:2011mr, Alloul:2013raa} (doublet).}
Its collider phenomenology has been investigated in various earlier works~\cite{Akeroyd:2005gt, Akeroyd:2009hb, Akeroyd:2010ip, Akeroyd:2011zza, Aoki:2011pz, Han:2007bk, Kanemura:2014goa, Kanemura:2014ipa, Vega:1989tt, Han:2003wu, Maalampi:2002vx, Chiang:2012dk, Chiang:2012cn, Shen:2014rpa, Chiang:2015rva, Chiang:2015amq, Chiang:2021lsx, Godfrey:2010qb, Englert:2013wga, Duarte:2022xpm, Jia:2024wqi, Sugiyama:2012yw, Nomura:2017abh, Gunion:1995mq, Rentala:2011mr, Alloul:2013raa, FileviezPerez:2008jbu, Han:2005nk, Melfo:2011nx, Das:2024kyk, Chakraborti:2023mya, Ashanujjaman:2022ofg, Ashanujjaman:2021txz, Ashanujjaman:2025puu, Mukhopadhyaya:2005vf, Dev:2019hev, Ghosh:2025gdx, Chiang:2013rua}. 
The indirect signals in the Higgs precision measurements have also been studied~\cite{Chiang:2018xpl, Chiang:2017vvo, Akeroyd:2012ms, Aoki:2012yt, Aoki:2012jj, Kanemura:2013mc, Chiang:2013rua, Chiang:2014bia, Englert:2013zpa, Hartling:2014zca, Hartling:2014aga, Das:2018vkv}.

Properties of the doubly charged Higgs bosons closely depend on the size of the vacuum expectation value (VEV) of the triplet field. 
Generally, in models such as the HTM, it is severely constrained by electroweak precision measurements because the triplet VEV breaks the custodial symmetry in the Higgs potential~\cite{Sikivie:1980hm}. 
On the other hand, in the GM model, the custodial symmetry is preserved at the tree level by assuming an alignment between the VEVs of the two triplet fields (of $Y = 0$ and $Y = 1$). 
It allows the triplet VEV to be relatively large compared with the HTM model and to enhance the three-point interaction among the doubly charged Higgs boson and the $W$ bosons.

Recently, the potential of high-energy same-sign muon colliders has been intensely studied~\cite{MuonCollider:2022nsa, Hamada:2022mua, Accettura:2023ked, InternationalMuonCollider:2025sys}. 
$\mu$TRISTAN at J-PARC, for example, is one of the options, where a low-emittance $\mu^+$ beam is achieved by using ultra-cold muons, and the center-of-mass energy of $\mu^+ \mu^+$ beams can reach the TeV scale~\cite{Hamada:2022mua}. 
The possibility of the same-sign electrons or positrons is also discussed~\cite{CLICPhysicsWorkingGroup:2004qvu}.
This situation strongly motivates studies on new physics searches in high-energy same-sign lepton colliders.

For the above reason, this paper investigates the search for doubly charged Higgs bosons at the same-sign lepton colliders. 
We consider the single production via the $W$ boson fusion: $\ell^+ \ell^+ \to H_5^{++} \bar{\nu}_\ell \bar{\nu}_\ell$.  
To enhance the production rate, we consider the GM model, where the doubly charged Higgs boson $H_5^{\pm\pm}$ is part of the quintet under the custodial symmetry of the model.  
The triplet VEV is set to be $\mathcal{O}(10)~\mathrm{GeV}$, so that $H_5^{++}$ predominantly decays into $W^+W^+$ rather than a pair of same-sign leptons because a larger triplet VEV leads to a smaller Yukawa coupling between the triplet and the charged leptons to maintain the size of the neutrino masses.
Thus, the signal is the $W$ boson scattering distorted by the resonant production of $H_5^{\pm\pm}$. 
Although the same signal is also expected in hadron colliders, lepton colliders provide a more effective platform for measuring the signal due to their lower QCD background.

We focus on the leptonic decays of the produced $W$ bosons to avoid the loss of information on the electric charge of particles. 
In this case, we cannot use the transverse mass of the decay product $\ell^\pm \ell^\pm \cancel{E}$ to obtain the mass of $H_5^{\pm\pm}$ because neutrinos radiated from the initial-state charged leptons constitute another source of missing energy. 
Instead, we propose to make use of the invariant mass of the charged leptons. 
By using the analytical formula for the invariant mass distribution, we can determine the best-fit value of the doubly charged Higgs boson mass. 
We demonstrate that this method is effective in obtaining information on the $H_5^{\pm\pm}$ mass, provided they are not too heavy.

We also discuss pair productions of the doubly charged Higgs bosons at opposite-sign lepton colliders.  We compare them with the single production at same-sign colliders.

This paper is organized as follows. 
In the next section, we introduce the GM model and give relevant formulas. 
In Sec.~\ref{sec: Hpp_at_lepton_colliders}, we discuss the $\ell^+ \ell^+ \to H_5^{++} \bar{\nu}_\ell \bar{\nu}_\ell \to W^+W^+\bar{\nu}_\ell \bar{\nu}_\ell$ process, as well as the SM background processes. 
In Sec.~\ref{sec: finv}, we discuss the discovery reach of the signal and the method to extract the mass of $H_5^{\pm\pm}$ in the leptonic decay mode of the $W$ bosons. 
We propose using the dilepton invariant mass for mass determination. 
We perform a numerical simulation and demonstrate that this method is effective unless the mass of $H_5^{\pm\pm}$ is too heavy.
The advantage of the single production compared to pair production processes at $\ell^+ \ell^-$ colliders is discussed in Sec.~\ref{sec: opposite-sign_colliders}. 
The conclusion is presented in Sec.~\ref{sec: conclusion}. 
In Appendix~\ref{app: finv}, we show the detailed derivation of the invariant mass distribution of the same-sign leptons in the decay of $H_5^{++}$.

\section{Georgi-Machacek Model}
\label{sec: GM_model}

The GM model features an extended Higgs sector comprising an isospin doublet $\phi$ with hypercharge $Y=1/2$ and isospin triplets $\chi$ and $\xi$ with hypercharges $Y=1$ and $Y=0$, respectively. 
We write them in the $SU(2)_L\times SU(2)_R$ covariant form \footnote{The convention for the negatively charged fields are given by $\phi^- = \left(\phi^+\right)^*$, $\chi^- = \left(\chi^+\right)^*$, $\xi^- = \left(\xi^+\right)^*$, and $\chi^{--} = \left(\chi^{++}\right)^*$.}: 
\begin{align}
\label{eq: Lagrangian fields}
\Phi  =
    \begin{pmatrix}
        \left(\phi^0\right)^* & \phi^+\\
        -\phi^- & \phi^0
    \end{pmatrix}, \quad 
\Delta  = 
    \begin{pmatrix}
    \left(\chi^0\right)^* & \xi^+ & \chi^{++}\\
    -\chi^- & \xi^0 & \chi^+\\
    \chi^{--} & -\xi^- & \chi^0\end{pmatrix}. 
\end{align}
Under global $SU(2)_L \times SU(2)_R$ transformations, $\Phi$ and $\Delta$ are transformed as $\Phi \to U_L \Phi U_R^\dagger$ and $\Delta \to U_L \Delta U_R^\dagger$, respectively, with $U_{L,R} = \exp(i\theta_{L,R}t^a)$, where $t^a$ ($a=1,2,3$) are generators of the $SU(2)$ Lie algebra, and $\theta_{L,R}$ are parameters of the transformations. 
The matrix forms of $t^a$ are given by $\sigma^a/2$ and $T^a$ for $\Phi$ and $\Delta$, respectively, where $\sigma^a$ are the Pauli matrices, and $T^a$ are defined as 
\begin{align}
    T^1 = \frac{1}{\sqrt{2}}
    \begin{pmatrix}
       0 & 1 & 0\\
       1 & 0 & 1\\
       0 & 1 & 0
    \end{pmatrix},\quad
    T^2 = \frac{1}{\sqrt{2}}
    \begin{pmatrix}
       0 & -i & 0\\
       i & 0 & -i\\
       0 & i & 0
    \end{pmatrix},\quad
    T^3 = 
    \begin{pmatrix}
       1 & 0 & 0\\
       0 & 0 & 0\\
       0 & 0 & -1
    \end{pmatrix}. 
\end{align}

The most general Higgs potential invariant under the global $SU(2)_L \times SU(2)_R \times U(1)_Y$ symmetry is given by
\begin{align}
V = & \frac{1}{2}m_1^2\mathrm{tr}[\Phi^{\dagger}\Phi] 
    + \frac{1}{2}m_2^2\mathrm{tr}[\Delta^{\dagger}\Delta] 
    + \lambda_1\left(\mathrm{tr}[\Phi^{\dagger}\Phi]\right)^2 
    + \lambda_2\left(\mathrm{tr}[\Delta^{\dagger}\Delta]\right)^2
\nonumber \\[5pt]
    & + \lambda_3\mathrm{tr}\Bigl[\left(\Delta^{\dagger}\Delta\right)^2\Bigr] 
    + \lambda_4\mathrm{tr}\bigl[\Phi^{\dagger}\Phi \bigr]\mathrm{tr}\bigl[\Delta^{\dagger}\Delta \bigr] 
    + \lambda_5 \mathrm{tr} \biggl[\Phi^{\dagger}\frac{\sigma^a}{2}\Phi\frac{\sigma^b}{2} \biggr]
        \mathrm{tr}\Bigl[\Delta^{\dagger}T^a\Delta T^b \Bigr]
\nonumber \\[5pt]
    & + \mu_1 \mathrm{tr}\biggl[\Phi^{\dagger}\frac{\sigma^a}{2}\Phi\frac{\sigma^b}{2} \biggr]
        \left(P^{\dagger}\Delta P\right)_{ab} 
    + \mu_2 \mathrm{tr}\Bigl[\Delta^{\dagger}T^a\Delta T^a\Bigr]
        \left(P^{\dagger}\Delta P\right)_{ab},
\end{align}
where the matrix $P$ is given by
\begin{align}
    P = \frac{1}{\sqrt{2}}
    \begin{pmatrix}
        -1 & i & 0\\
        0 & 0 & \sqrt{2}\\
        1 & i & 0
    \end{pmatrix}, 
\end{align}
which relates $T^a$ to the adjoint representation of $t^a$: $P^\dagger T^a P = \mathrm{ad}(t^a)$.

The neutral scalar fields are parametrized by $\phi^0=\frac{1}{\sqrt{2}}(\phi_r+i\phi_i)$, $\chi^0 = \frac{1}{\sqrt{2}}(\chi_r+i\chi_i)$, and $\xi^0 = \xi_r$. 
They acquire the following VEVs
\begin{align}
    \left<\phi_r\right> = v_{\Phi}, \quad \frac{1}{\sqrt{2}}\left<\chi_r\right> = \left<\xi_r\right> = v_{\Delta},  
\end{align}
and the others are zero. 
These VEVs break the electroweak symmetry and satisfy $v^2 = v_{\Phi}^2 + 8v_{\Delta}^2 \approx (246~\mathrm{GeV})^2$. 
In addition, the global $SU(2)_L \times SU(2)_R$ symmetry is spontaneously broken to the custodial symmetry $SU(2)_V$ in the Higgs potential. 
For later convenience, we define parameters $\beta$, $M_1^2$, and $M_2^2$ as 
\begin{equation}
\tan\beta = \frac{v_{\Phi}}{2\sqrt{2}v_{\Delta}},
   \quad M_1^2 = -\frac{v\mu_1}{\sqrt{2}c_\beta},
        \quad M_2^2 = -3\sqrt{2}v\mu_2 c_\beta, 
\end{equation}
where $c_\beta = \cos \beta$ and $s_\beta = \sin \beta$ are used throughout the paper.

The stationary condition requires
\begin{equation}
\frac{\partial V}{\partial \phi_r}=0, \quad 
    \frac{\partial V}{\partial \chi_r}=0, \quad 
        \frac{\partial V}{\partial \xi_r}=0, 
\end{equation}
at the vacuum, with the latter two conditions being equivalent.  We thus obtain two independent equations:
\begin{align}
\begin{split}
& m_1^2 = -4\lambda_1v_{\Phi}^2 -6\lambda_4v_{\Delta}^2 - 3\lambda_5v_{\Delta}^2 - \frac{3}{2}\mu_1v_{\Delta}, \\
& m_2^2 = -12\lambda_2v_{\Delta}^2 - 4\lambda_3v_{\Delta}^2 - 2\lambda_4v_{\Phi}^2 -\lambda_5v_{\Phi}^2 -\mu_1\frac{v_{\Phi}^2}{4v_{\Delta}} - 6\mu_2v_{\Delta}.
\end{split}
\end{align}

Since the Higgs potential possesses the custodial symmetry at the tree level, 
it is convenient to classify the scalar fields into irreducible representations of $SU(2)_V$: a quintet (or 5-plet) $H_5$, a triplet $H_3$, and two singlets $H_1$ and $h$.  
Each component of the multiplets and singlet fields is defined in terms of the fields in Eq.~\eqref{eq: Lagrangian fields} as 
\begin{align}
\begin{split}
& H_5^{\pm\pm}=\chi^{\pm\pm}, 
    \quad H_5^{\pm}=\frac{1}{\sqrt{2}}\left(\chi^{\pm}-\xi^{\pm}\right), 
        \quad H_5^0=\sqrt{\frac{1}{3}} \chi_r-\sqrt{\frac{2}{3}} \xi_r, 
\\
& H_3^{\pm}=- c_\beta \phi^{\pm}+ \frac{s_\beta}{\sqrt{2}}\left(\chi^{\pm}+\xi^{\pm}\right), 
    \quad H_3^0=-c_\beta \phi_i+ s_\beta \chi_i,
\\
& h= c_\alpha \phi_r-\frac{s_\alpha}{\sqrt{3}}\left(\sqrt{2} \chi_r+\xi_r\right), \quad H_1= s_\alpha \phi_r+\frac{c_\alpha}{\sqrt{3}}\left(\sqrt{2} \chi_r+\xi_r\right),     
\end{split}
\end{align}
where $s_\alpha = \sin \alpha$, $c_\alpha = \cos \alpha$, and $\alpha$ satisfies $\tan 2\alpha = 2M^2_{12}/(M^2_{22}-M^2_{11})$ with 
\begin{align}
\begin{split}
    &
    M^2_{11} = 8\lambda_1 v^2 s_\beta^2,
        \quad M^2_{22} = (3\lambda_2 + \lambda_3) v^2 c_\beta^2 + M^2_1s_\beta^2 -\frac{1}{2}M^2_2, 
    \\
    &
    M^2_{12} = \sqrt{\frac{3}{2}} \Bigl((2\lambda_4+\lambda_5)v^2-M^2_1 \Bigr)s_\beta c_\beta.
\end{split}
\end{align}
The mass formulas of the quintet, the triplet, $H_1$, and $h$ are given by
\begin{align}
\begin{split}
& m_{H_5}^2 
=
\biggl(M_1^2-\frac{3}{2} \lambda_5 v^2 \biggr) s_\beta^2 
    +\lambda_3 v^2 c_\beta^2 +M_2^2,
\\
& m_{H_3}^2 
= 
M_1^2-\frac{1}{2} \lambda_5 v^2,
\\
& m_{H_1}^2 
=
M^2_{11} s_\alpha^2 + M^2_{22} c_\alpha^2 + 2M^2_{12} s_\alpha c_\alpha,
\\
& m_h^2 
= 
M^2_{11} c_\alpha^2 +M^2_{22}s_\alpha^2 -2M^2_{12} s_\alpha c_\alpha, 
\end{split}
\end{align}
respectively. In the following, $h$ is identified as the observed 125-GeV SM-like Higgs boson. 
We neglect small mass splittings of $\mathcal{O}(1)~\mathrm{GeV}$ among the multiplets due to one-loop radiative corrections~\cite{Blasi:2017xmc}.

Before closing this section, we briefly discuss the experimental constraints on the model.
Focusing on $H_5^{\pm\pm}$ in this analysis, we consider only constraints on the quintet states. 
The phenomenology of the quintet is described purely by two parameters: $m_{H_5}$ and $\beta$. 
We use the summary plot in Ref.~\cite{ATLAS:2022hnp} to extract the experimental constraint from the searches for heavy diboson resonances in semileptonic states~\cite{ATLAS:2020fry}, heavy resonances decaying into a pair of $Z$ bosons~\cite{ATLAS:2020tlo}, and the multilepton final states via the vector bosons coming from the decays of the pair-produced $H_5^{++} H_5^{--}/H_5^{\pm\pm} H_5^\mp$~\cite{ATLAS:2021jol}. 
In addition, we refer to Ref.~\cite{ATLAS:2024txt} for the constraint from the combination of searches for the single production of the charged Higgs bosons with the bosonic decays: $H_5^\pm \to W^\pm Z$~\cite{ATLAS:2022zuc} and $H_5^{\pm\pm} \to W^\pm W^\pm$~\cite{ATLAS:2023sua}.
The $H^{++}_5 H^{--}_5 \to 4W$ search has excluded the region $m_{H_5} < 350~\mathrm{GeV}$, assuming $\mathrm{Br}(H_5^{\pm\pm} \to W^\pm W^\pm) = 100~\%$, {\it i.e.}, $v_\Delta \gtrsim 10^{-3}$ GeV ($\cos \beta \gtrsim 10^{-5}$).
For heavier mass $m_{H_5} > 350~\mathrm{GeV}$, the upper bound on $v_\Delta$ depends on $m_{H_5}$; for example, $v_\Delta \lesssim 22~\mathrm{GeV}$ ($\cos \beta \lesssim 0.25$) for $m_{H_5} = 400~\mathrm{GeV}$.
The complete constraint in $m_{H_5}^{}$-$v_\Delta$ ($m_{H_5}^{}$-$\cos\beta$) plane is shown in Fig.~\ref{fig:Contours_for_Significance}.\footnote{We note that the angle $\theta_H$ used in Refs.~\cite{ATLAS:2022hnp,ATLAS:2024txt} is related to $\beta$ through $\sin \theta_H = \cos \beta = 2\sqrt{2}v_\Delta/v$.}

\section{Doubly charged Higgs boson at same-sign lepton colliders}
\label{sec: Hpp_at_lepton_colliders}

In this section, we discuss the single production process of the doubly charged Higgs boson at same-sign lepton colliders. 
In the following analyses, we assume that the initial state is $\mu^+ \mu^+$ as an explicit example due to its experimental advantages in reaching high-energy frontier: (i) less synchrotron radiation than electrons/positrons in acceleration, and (ii) the established technology to produce a low-emittance $\mu^+$ beam by using the ultra-cold muons.

From the theoretical point of view, there is no significant difference among the beam options because the signal and background processes are dominantly generated by the gauge interaction. 
Thus, our result is also valid for the processes at lepton colliders with other beam options: $e^+ e^+$, $e^+ \mu^+$, and their charge conjugations.

\subsection{Setup for numerical simulations}

Here, we explain the setup for numerical evaluations in the rest of this paper. 
In view of the interest in $H_5^{\pm\pm}$, we assume the other additional Higgs bosons $H_3$ and $H_1$ are sufficiently heavier than $H_5$ and neglect their contributions.\footnote{In Sec.~\ref{sec: opposite-sign_colliders}, we switch on the contribution of $H_3$ in studying a pair production via the vector boson fusion because it is necessary to cancel the quadratic dependence on the beam energy due to longitudinal vector bosons~\cite{Lee:1977eg}.} 
Such a mass hierarchy can be realized by an appropriate choice of the model parameters~\cite{Logan:2017jpr,Chen:2022zsh}.
Also, we choose the value of the mixing angle $\alpha$ so that the couplings of $h$ coincide with those of the SM Higgs boson at the tree level. 
This value also depends on the choice of $v_\Delta$. 
See Fig.~1 of Ref.~\cite{Chen:2022zsh} for the details. 
With these assumptions, the new physics effects manifest only through the quintet Higgs bosons at lower energies.

For numerical simulations, we use MadGraph5\_aMC@NLO (MG5)~\cite{Alwall:2014hca}. 
For the SM background processes, we use the default model file in MG5. 
In evaluating processes in the GM model, we use a public UFO file.\footnote{\url{https://cp3.irmp.ucl.ac.be/projects/feynrules/wiki/GeorgiMachacekModel}}
We note that these models use different input values for the fine structure constant $\alpha_\mathrm{EM}$ at the $Z$ pole due to different renormalization schemes. 
In the SM, $\alpha^{-1}_\mathrm{EM} = 132.507$ is employed, while $\alpha^{-1}_\mathrm{EM} = 127.9$ in the GM model. 
For a correct comparison between the signal and background processes, 
we use $\alpha^{-1}_\mathrm{EM} = 132.507$ in both models. 
To make plots, we used MadAnalysis~\cite{Conte:2012fm}.

\subsection{The single production of $H_5^{++}$}

Here, we consider the single production at $\mu^+\mu^+$ colliders, $\mu^+ \mu^+ \to H_5^{++} \bar{\nu}_\mu \bar{\nu}_\mu$ via the $W^+$ boson fusion. 
The process is induced by the $W^+W^+H_5^{--}$ coupling, which is proportional to $v_\Delta$. 
Thus, the cross section of the process is determined by two new physics parameters, $v_\Delta$ and $m_{H_5}$, and the beam energy, $\sqrt{s}$.

\begin{figure}[t]
    \subfigure[\ $\sigma$ vs. $\sqrt{s}$]{
         \includegraphics[width=0.46\textwidth]{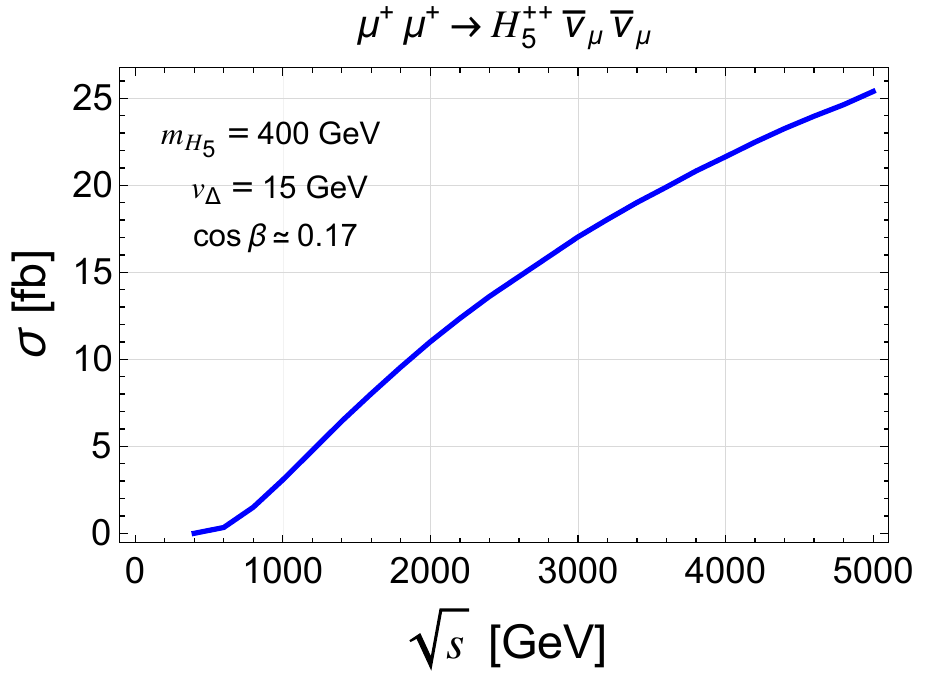}
         \label{fig: sigma_vs_s}
         }
     \subfigure[\ $\sigma$ vs. $m_{H_5}$]{
         \includegraphics[width=0.46\textwidth]{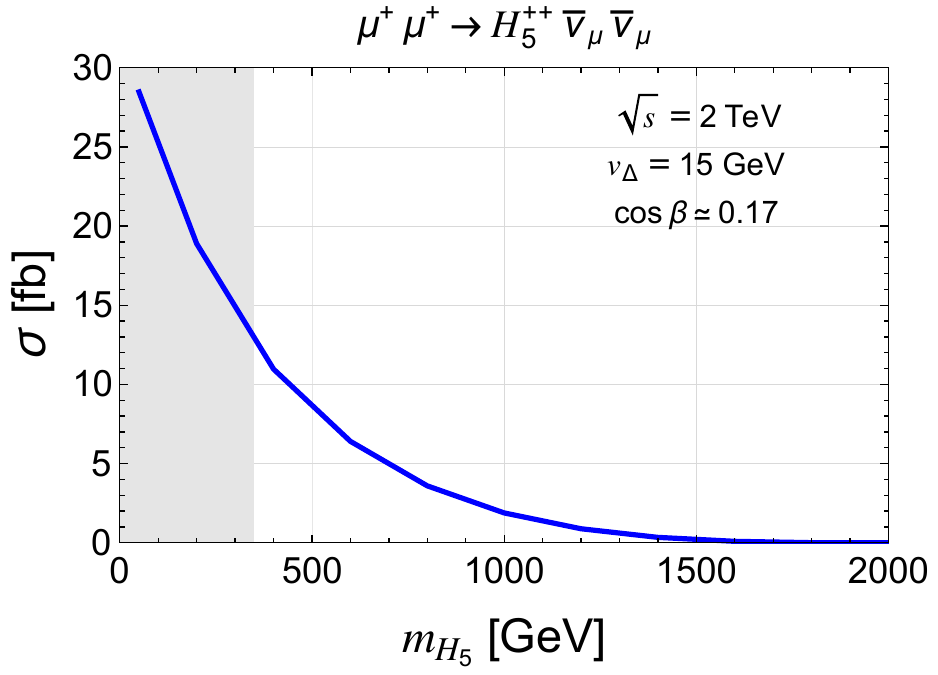}
         \label{fig: sigma_vs_m}
         }
     \subfigure[\ $\sigma$ vs. $v_{\Delta}$]{
         \includegraphics[width=0.46\textwidth]{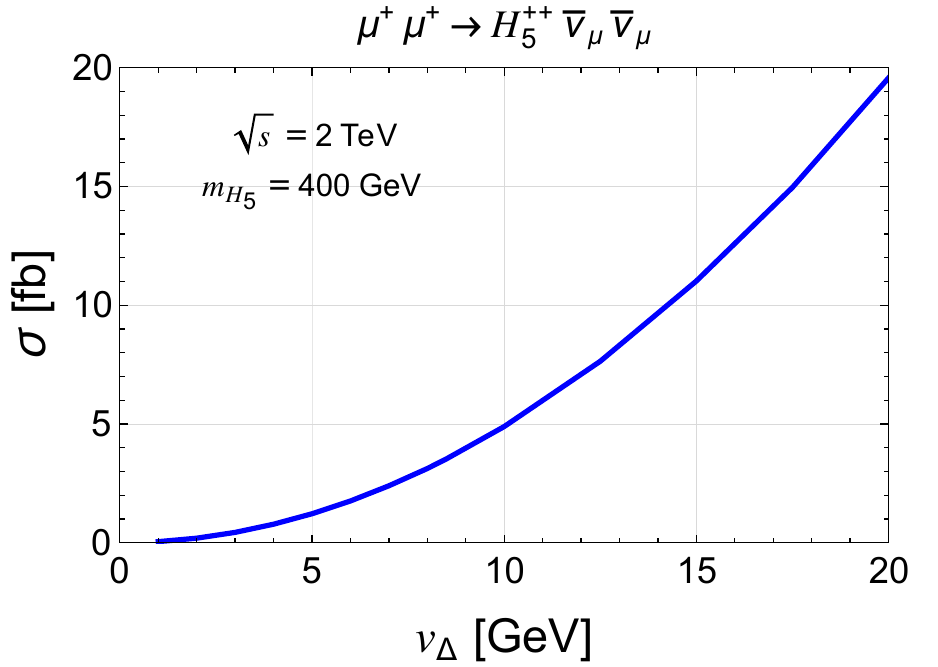}
         \label{fig: sigma_vs_v}
         }
    \caption{The dependence of the cross section $\sigma$ on (a) $\sqrt{s}$, (b) $m_{H_5}$, and (c) $v_\Delta$. 
    In each figure, the parameters not on the horizontal axis are fixed at $\sqrt{s}=2~\mathrm{TeV}$, $m_{H_5}=400~\mathrm{GeV}$, and $v_{\Delta}=15~\mathrm{GeV}$ ($\cos \beta \simeq 0.17$). The gray regions are excluded by the experimental constraints discussed in Sec.~\ref{sec: GM_model}.}
    \label{fig: signal_xsec}
\end{figure}

We show the dependence of the cross section $\sigma$ on each of these three parameters in Fig.~\ref{fig: signal_xsec}. 
In Fig.~\ref{fig: sigma_vs_s}, the beam energy dependence is shown with $m_{H_5} = 400~\mathrm{GeV}$ and $v_\Delta = 15~\mathrm{GeV}$ ($\cos \beta \simeq 0.17$). 
Above the threshold, $\sqrt{s}\simeq m_{H_5}$, $\sigma$ increases logarithmically with the beam energy. 
In Fig.~\ref{fig: sigma_vs_m}, $\sigma$ is shown in the range $0 < m_{H_5} <2~\mathrm{TeV}$ with $\sqrt{s} = 2~\mathrm{TeV}$ and $v_\Delta = 15~\mathrm{GeV}$.  
In Fig.~\ref{fig: sigma_vs_v}, $\sigma$ is shown as a function of $v_\Delta$ with $m_{H_5} = 400~\mathrm{GeV}$ and $\sqrt{s} = 2~\mathrm{TeV}$.  
Since the $W^+W^+H^{--}$ coupling is proportional to $v_\Delta$, $\sigma$ increases quadratically with $v_\Delta$. 
In Figs.~\ref{fig: sigma_vs_m} and~\ref{fig: sigma_vs_v}, the gray regions are excluded by the experimental constraints discussed in the previous section.

\subsection{The signal and background processes}

\begin{figure}[t]
    \begin{center}
    \includegraphics[width=1.0\textwidth]{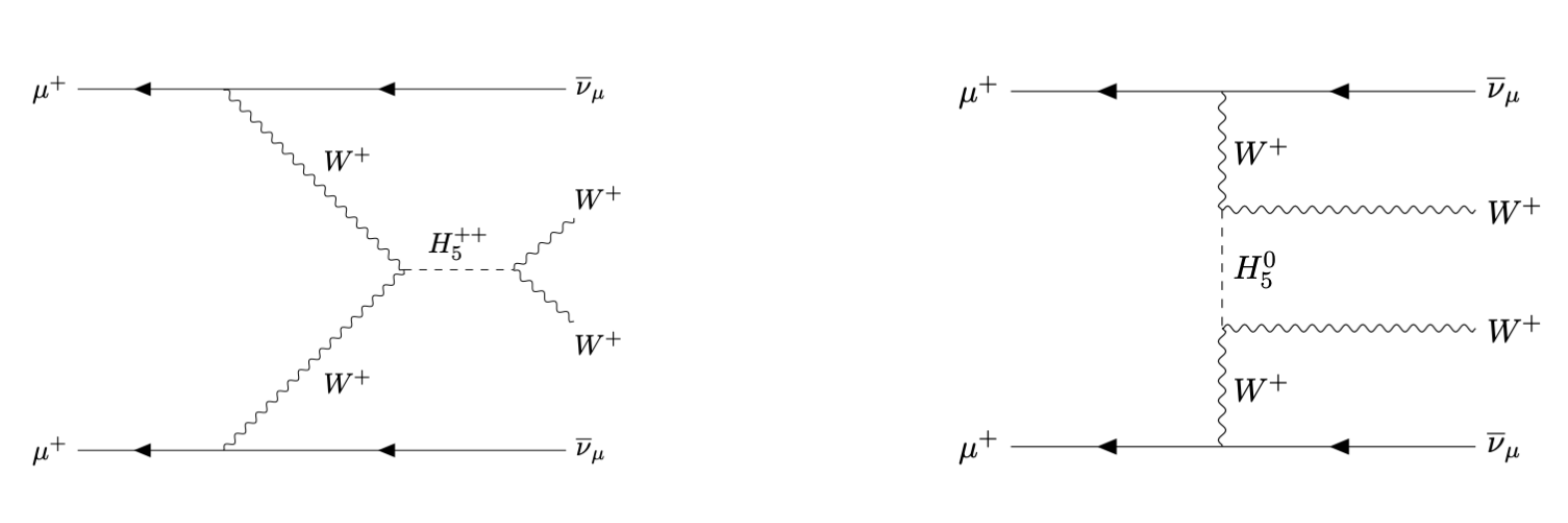}
    \caption{Feynman diagrams of new physics contributions for the signal process.}
    \label{fig: Feynman_diagrams}
    \end{center}
\end{figure}

We assume that $v_\Delta$ is sufficiently large, {\it i.e.}, $v_\Delta \gtrsim 10^{-3}~\mathrm{GeV}$ ($\cos \beta \gtrsim 10^{-5}$) so that the produced $H_5^{++}$ decays into a pair of $W$ bosons virtually 100 percent~\cite{Han:2005nk, Melfo:2011nx}. 
The signal is $\mu^+ \mu^+ \to W^+ W^+ \bar{\nu}_\mu \bar{\nu}_\mu$. 
In this process, there is another new physics contribution via t-channel $H_5^0$ mediation. 
However, the dominant new physics effect comes from the resonant production of $H_5^{++}$. 
The Feynman diagrams of the two processes are shown in Fig.~\ref{fig: Feynman_diagrams}.

\begin{figure}[t]
    \begin{center}
    \subfigure[\ $\sigma$ vs. $\sqrt{s}$ ]{
         \includegraphics[width=0.46\textwidth]{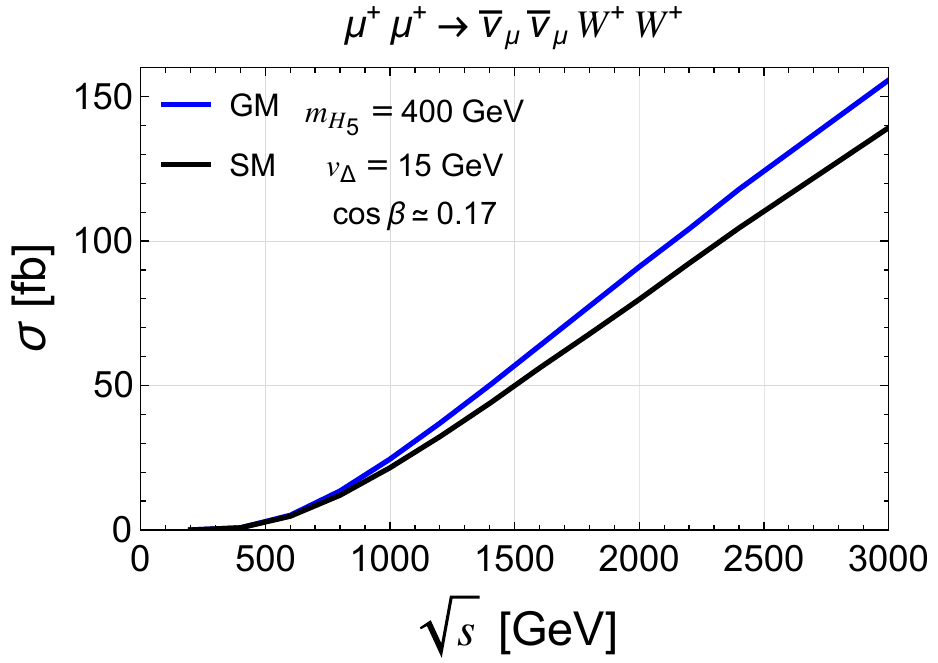}
         \label{fig: sigma_vs_s_2}
         }
     \subfigure[\ $\sigma$ vs. $m_{H_5}$ ]{
         \includegraphics[width=0.46\textwidth]{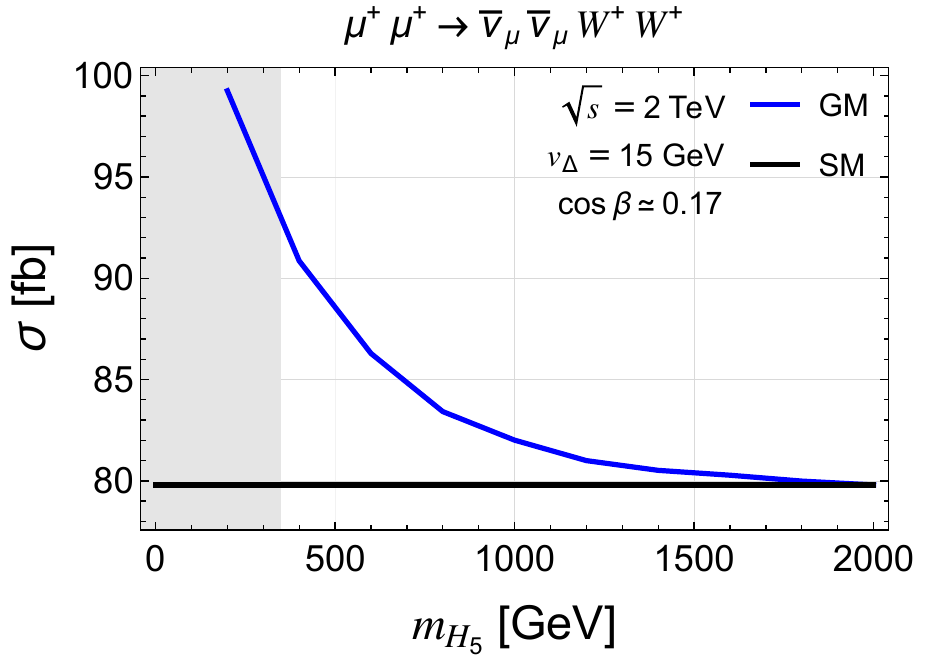}
         \label{fig: sigma_vs_m_2}
         }
     \subfigure[\ $\sigma$ vs. $v_{\Delta}$ ]{
         \includegraphics[width=0.46\textwidth]{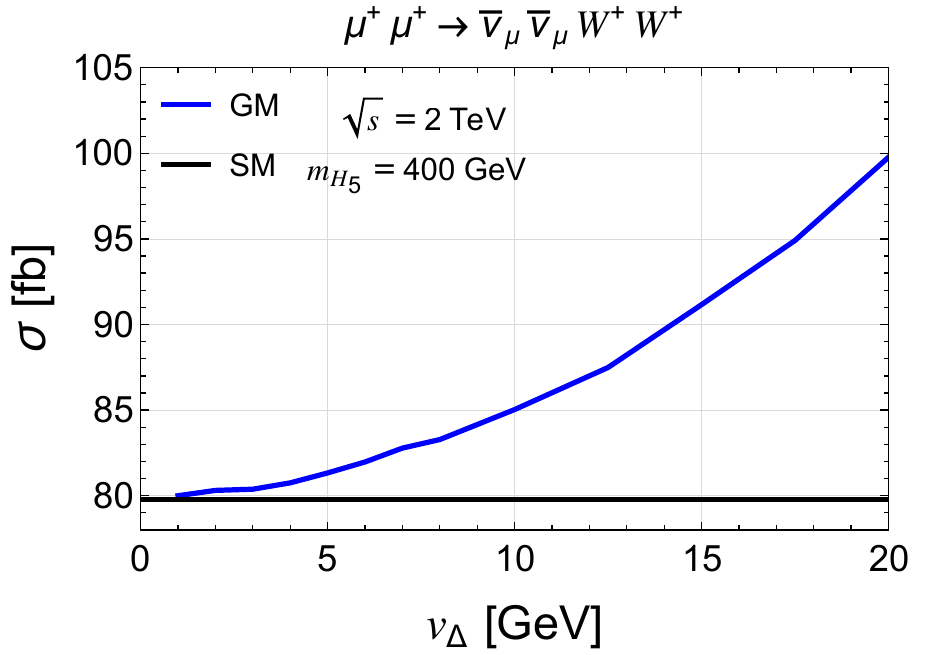}
         \label{fig: sigma_vs_v_2}
         }
    \caption{Comparisons between the cross section of $\mu^+ \mu^+ \to \bar{\nu}_\mu \bar{\nu}_\mu W^+W^+$ in the SM and the GM model as functions of (a) $\sqrt{s}$, (b) $m_{H_5}$, and (c) $v_\Delta$. 
    In each figure, the parameters not in the horizontal axis are fixed at $\sqrt{s}=2~\mathrm{TeV}$, $m_{H_5}=400~\mathrm{GeV}$, and $v_{\Delta}=15~\mathrm{GeV}$ ($\cos \beta \simeq 0.17$).}
    \label{fig: signal_vs_BG}
    \end{center}
\end{figure}

The main background process is the $W$ boson scattering mediated by the SM particles. 
In Fig.~\ref{fig: signal_vs_BG}, we compare the cross section of $\mu^+ \mu^+ \to \bar{\nu}_\mu \bar{\nu}_\mu W^+ W^+$ in the SM and the GM model as functions of (a) $\sqrt{s}$, (b) $m_{H_5}$, and (c) $v_\Delta$. 
In each figure, the black and blue curves represent the cross section in the SM and the GM model, respectively. 
In Figs.~\ref{fig: sigma_vs_m_2} and \ref{fig: sigma_vs_v_2}, the black curve is a constant because the cross section in the SM is independent of the new physics parameters. 
A general trend is that the new physics contribution is larger for larger $\sqrt{s}$, lower $m_{H_5}$, and larger $v_\Delta$. 
For example, with $\sqrt{s} = 2~\mathrm{TeV}$, $m_{H_5} = 400~\mathrm{GeV}$, and $v_\Delta = 15~\mathrm{GeV}$, the cross section is $\sigma \simeq 92~\mathrm{fb}$ in the GM model, while $\sigma = 80~\mathrm{fb}$ in the SM, with the expected number of events about $15~\%$ larger in the former.

In the above discussions, we have assumed unpolarized $\mu^+$ beams. 
Since the $W$ boson couples to only $\mu^+$ with the right-handed chirality, the cross section will be multiplied by the factor $(1+P_{\mu_1^+})(1+P_{\mu_2^+})$ in both SM and GM model if we use the polarized $\mu^+$ beams, where $P_{\mu_i^+}$ ($i=1,2$) are the polarization rate ranging from $-1$ to $1$~\cite{Ahmed:2022wlz}. 
For example, with $80\%$ right-handed polarized $\mu^+$ beams, the cross section is $1.8^2 \simeq 3.2$ times larger than the unpolarized cross section.

\section{The discovery reach and mass determination in the same-sign lepton signal}
\label{sec: finv}

In this section, we discuss the discovery reach of the signal and a way to determine the mass of $H_5^{++}$.
To avoid the loss of information on the electric charge of particles, we consider the leptonic decay of the $W$ boson: $W^+ \to \ell^+ \nu_\ell$ ($\ell = e$ or $\mu$).
Then the final state of the signal is $\ell^+ \ell^{\prime +} \nu_{\ell} \nu_{\ell^\prime} \bar{\nu}_\mu \bar{\nu}_\mu$, {\it i.e.}, a pair of same-sign leptons and missing energy.

To reduce the background processes involving muons coming from the initial state and the invisible decay of the $Z$ boson, we consider the case of $\ell = \ell^\prime = e$ in the following. 
In addition, for the accurate selection of the signal events, we employ the following kinematical cuts for the phase space~\cite{ATLAS-CONF-2023-023};
\begin{equation}
\label{eq: kinematica_cut}
p_T^{e^+} > 27~\text{GeV}, 
    \quad \eta_{e^+} < 1.37,
        \quad M > 20~\text{GeV},
\end{equation}
where $p_T^{e^+}$ and $\eta_{e^+}$ are the transverse momentum and the rapidity of $e^+$, respectively, and $M$ is the invariant mass of the positron pair. 

\begin{table}[t]
\centering
\renewcommand{\arraystretch}{1.2}
\begin{tabular*}{0.8\textwidth}{@{\extracolsep{\fill}}c| c c c c c}
\toprule
$\sqrt{s}$ \textbackslash~$m_{H_5}$ 
& $400$~GeV & $500$~GeV & $600$~GeV  \\
\hline
$1$~TeV &  
1.6 (3.5) & 0.7 (1.6) & 0.0 (0.0)  \\
$2$~TeV & 
2.6 (5.7) & 2.0 (4.4) & 1.3 (2.8)  \\
$3$~TeV & 
2.4 (5.5) & 1.6 (3.5) & 1.4 (3.2)  \\
\hline
\end{tabular*}
\caption{The statistical significance of the same-sign lepton signal with \(v_{\Delta}=15\)~GeV for various beam energies $\sqrt{s}$ and the masses $m_{H_5}$. 
The values without and with parentheses represent the significance with the integrated luminosity $1000~\mathrm{fb^{-1}}$ and $5000~\mathrm{fb^{-1}}$, respectively.}
\label{ref: dicovery_reach}
\end{table}

In Table~\ref{ref: dicovery_reach}, we show the statistical significance of the signal for various masses and beam energies.
The values without (with) parenthesis are $S/\sqrt{S+B}$ with the integrated luminosity $\mathcal{L} = 1000~\mathrm{fb^{-1}}$ ($5000~\mathrm{fb^{-1}}$), where $B$ and $S$ are the number of the SM background events and the deviation of the expected number of events from the SM background, respectively. 

In the case of $\mathcal{L} = 1000~\mathrm{fb^{-1}}$, an even higher beam energy is required in order to exceed $S/\sqrt{S+B}=5$ although the evidence of the signal is expected to be observed, $i.e.$, $S/\sqrt{S+B}>3$, in the lower mass region ($m_{H_5} \lesssim 300$~GeV). 
On the other hand, with $\mathcal{L} = 5000~\mathrm{fb^{-1}}$, the signal is expected to be discoverable in some benchmark points on the table. 
For example, with $\sqrt{s} = 3$~TeV, the discovery reach is above $m_{H_5} = 400~\mathrm{GeV}$. 
Also, we expect to obtain the evidence of $H_5^{++}$ lighter than $600~\mathrm{GeV}$.

\begin{figure}[t]
    \centering
    \includegraphics[width=0.6\linewidth]{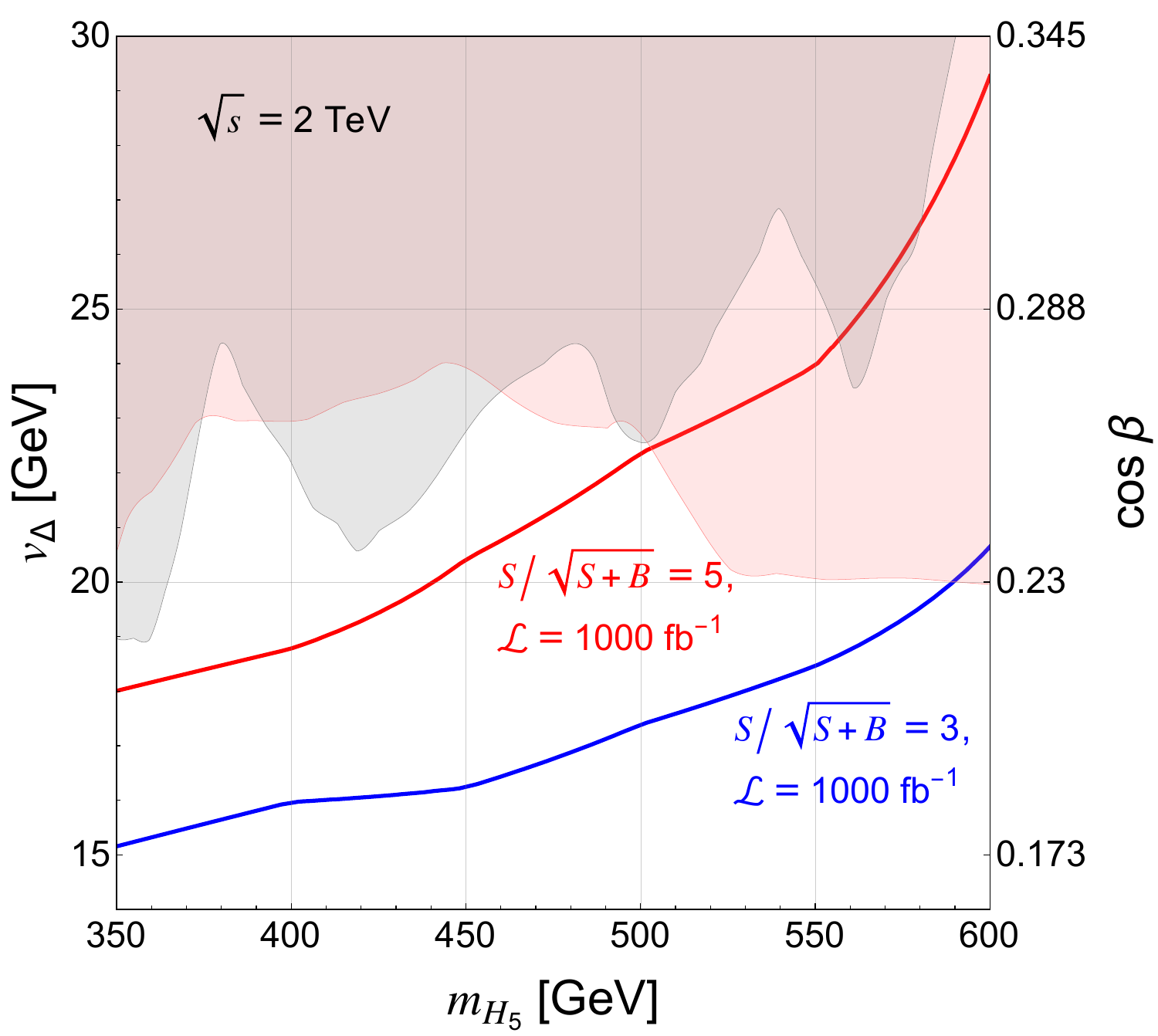}
    \caption{Contours for the statistical significance with $\sqrt{s} = 2~\mathrm{TeV}$ and $\mathcal{L} = 1000~\mathrm{fb^{-1}}$.}
    \label{fig:Contours_for_Significance}
\end{figure}

In Fig.~\ref{fig:Contours_for_Significance}, we show the statistical significance of the signal in the case of $\sqrt{s} = 2~\mathrm{TeV}$ and $\mathcal{L} = 1000~\mathrm{fb^{-1}}$ as a function of $m_{H_5}$ and $v_\Delta$ ($\cos \beta$), and the solid red and blue curves represent $S/\sqrt{S+B} = 5$ and 3, respectively.
The gray (pink) regions are excluded by Ref.~\cite{ATLAS:2022hnp} (\cite{ATLAS:2024txt}). 
In the regions $m_{H_5} \lesssim 500~\mathrm{GeV}$ and $m_{H_5} \lesssim 590~\mathrm{GeV}$, we can find parameter points that can be tested at the expected sensitivities of $5~\sigma$ and $3~\sigma$ levels, respectively. 
Larger integrated luminosities extend the testable areas furthermore.

For a comparison with the expected sensitivity of future hadron colliders, we refer to Ref.~\cite{Chiang:2021lsx}, which studies searches for the signature of same-sign $W$ bosons and forward jets arising from the production of doubly charged scalar bosons via the vector boson fusion process. 
Their analysis showed that future $pp$ colliders with a beam energy of $\sqrt{s} = 27~\mathrm{TeV}$ could reach $2.5 \sigma$ significance with $\mathcal{L} = 3000~\mathrm{fb^{-1}}$ for $m_{H_5} = 300~\mathrm{GeV}$ and $v_\Delta = 10~\mathrm{GeV}$. 
Assuming the simple scaling of $S \propto v_\Delta^2\mathcal{L}$ and $B \propto \mathcal{L}$, it corresponds to approximately $3.2 \sigma$ significance for $v_\Delta = 15~\mathrm{GeV}$ and $\mathcal{L} = 1000~\mathrm{fb^{-1}}$ with the same energy and mass. 
In contrast, our analysis results in $S/\sqrt{S+B} \simeq 3.2$ for the same input parameters and luminosity but $\sqrt{s} = 2~\mathrm{TeV}$. 
Thus, we expect that these two colliders would have roughly comparable sensitivities to the signal process.
Although more sophisticated background analysis for the same-sign lepton colliders is required for a precise comparison, high-energy same-sign lepton colliders may offer a useful and complementary platform for future searches of doubly charged Higgs bosons besides hadron colliders.

\begin{figure}[t]
    \subfigure[\ The transverse mass distribution]{
        \includegraphics[width=0.45\textwidth]{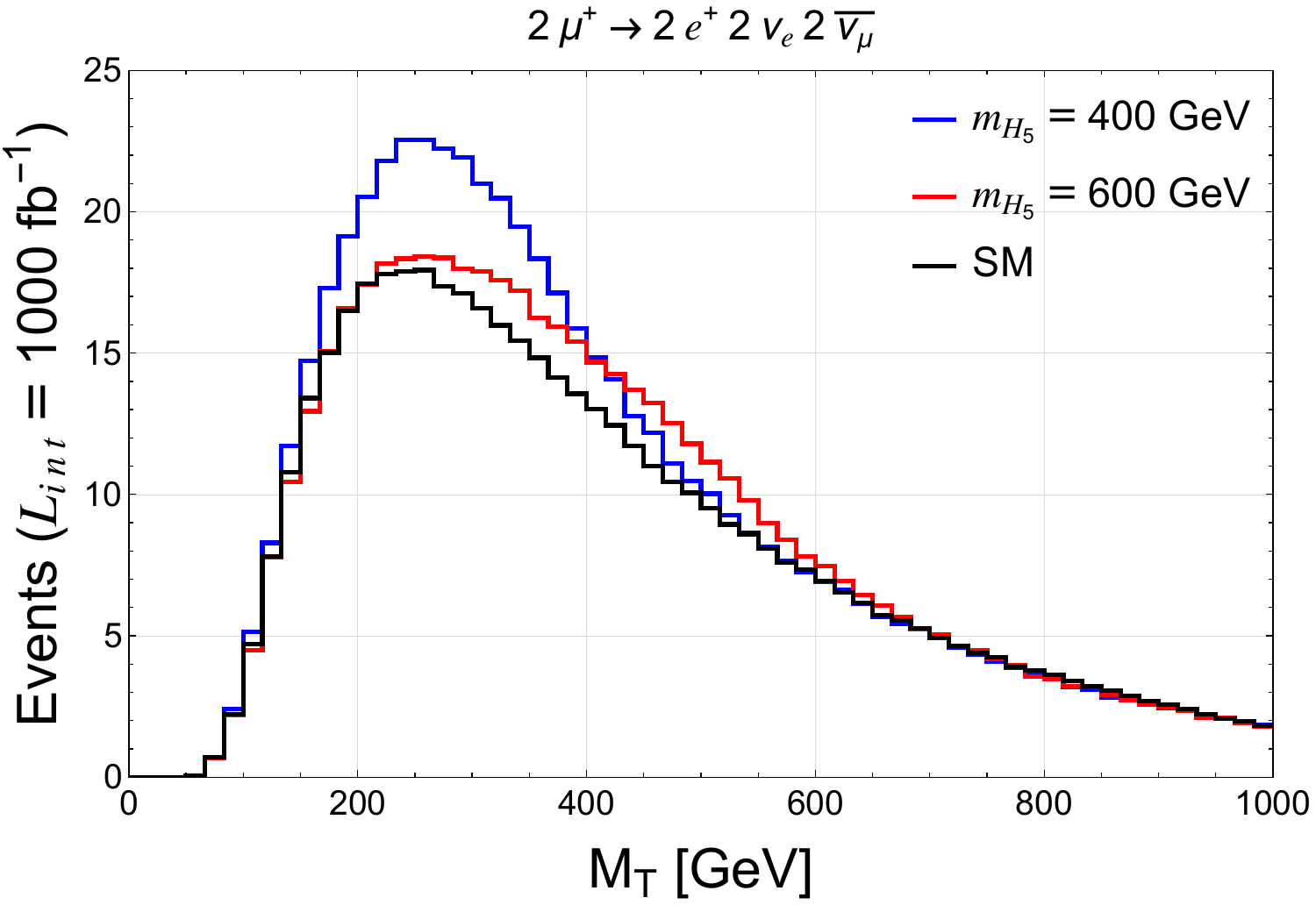}
        \label{fig: MT_dist}
    }
    \subfigure[\ The invariant mass distribution]{
        \includegraphics[width=0.45\textwidth]{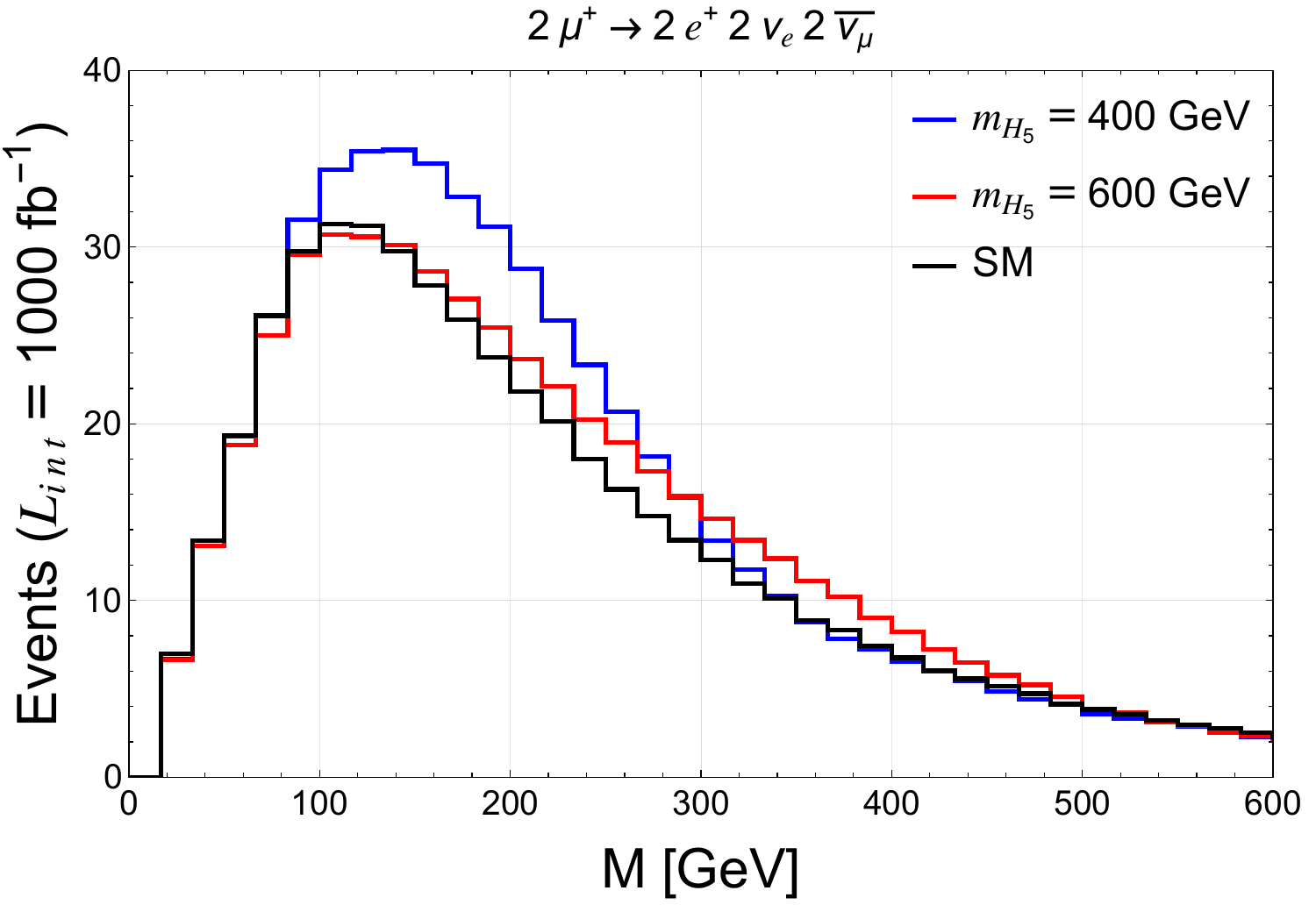}
        \label{fig: Minv_dist}
    }
    \caption{The distributions of (a) the transverse mass of $e^+ e^+ \cancel{E}$ and (b) the invariant mass of $e^+e^+$. 
    The red and blue curves are contour for $m_{H_5} = 600~\mathrm{GeV}$ and $m_{H_5} = 400~\mathrm{GeV}$, respectively.  Other parameters are fixed as $\sqrt{s} = 2~\mathrm{TeV}$ and $v_\Delta = 15~\mathrm{GeV}$ ($\cos \beta \simeq 0.17$).  The black curve represents the SM prediction.  For all curves, we use the kinematical cuts in Eq.~(\ref{eq: kinematica_cut}).}
    \label{fig: MT_Minv_dist}
\end{figure}

Next, we discuss a way to obtain the mass information of $H_5^{++}$ in the same-sign lepton signal.
In many cases, one can use the transverse mass of the decay product to measure the mass of particles when the decay product includes missing energy. 
The distribution of the transverse mass would have a Jacobian peak at the mass of the mother particle~\cite{Smith:1983aa, Barger:1983wf}. 
However, this does not apply to the current process. 
It is impossible to separate $\bar{\nu}_\mu$ radiated from the initial $\mu^+$ and $\nu_e$ in the decay product. 
We show the distribution of the transverse mass of $e^+ e^+\cancel{E}$ in Fig.~\ref{fig: MT_dist}. 
Although the distribution depends on the mass, there is no sharp peak at the mass of $H_5^{++}$.

Alternatively, we propose a method of using the invariant mass of the positrons. 
In Fig~\ref{fig: Minv_dist}, we show the invariant mass distribution of $e^+ e^+$.  
We perform a $\chi^2$ fit to obtain the best-fit value of $m_{H_5}$ by employing the analytic formula for the invariant mass distribution.

For the sake of simplicity, we consider the invariant mass distribution $f_\mathrm{inv}(M)$ in the decay of $H_5^{++}$, rather than evaluating the entire process. 
In this case, $f_\mathrm{inv}(M)$ approximately describes the distribution in the difference between the signals in the GM model and the SM background because the new physics effect is dominated by the on-shell $H_5^{++}$ production.
There are other contributions to the difference between the signal and the SM background: the interference terms and the t-channel diagram mediated by $H_5^0$ in Fig.~\ref{fig: Feynman_diagrams}. The rate of these effects in the deviation is larger for heavier $H_5^{\pm\pm}$, while the dependence on the beam energy is small in the region $1~\mathrm{TeV}<\sqrt{s}<3~\mathrm{TeV}$. Numerically, these effects account for about $1~\%$ of the total deviation for $m_{H_5} = 400~\mathrm{GeV}$.

The advantage of using the invariant mass in the fit rather than the transverse mass is as follows. 
Since $M$ is Lorentz invariant, its distribution is also Lorentz invariant.
Thus, in the $\chi^2$ fit, we can use a common $f_\mathrm{inv}(M)$ regardless the motion of $H_5^{++}$. 
On the other hand, the transverse mass is not Lorentz invariant. 
Thus, the motion of $H_5^{++}$ will distort the distribution.

A detailed derivation of $f_\mathrm{inv}(M)$ is presented in Appendix~\ref{app: finv}. 
Here, we show only the result:
\begin{align}
\label{eq: finv}
f_\mathrm{inv}(M) = N_\mathrm{ev} \frac{ 9(1+v) }{ 4 m_5}  \frac{ \sqrt{\omega} }{ 3 - 2v^2 + 3v^4 } F(v,\omega), 
\end{align}
where $N_\mathrm{ev}$ is the total number of events.  The parameters $v$ and $\omega$ are the speed of the $W$ bosons and the normalized invariant mass squared, respectively,
\begin{align}
v = \sqrt{ 1 - \frac{ 4m_W^2 }{ m_{H_5}^2 } },\quad \omega = \frac{ M^2 }{ M_\mathrm{max}^2 } 
    = \frac{ 4 M^2 }{ m_{H_5}^2 (1+v)^2 }, 
\end{align}
where $M_\mathrm{max}$ is the maximum value of the invariant mass. 
The definition of the function $F(v,\omega)$ is given in Appendix~\ref{app: finv}.

\begin{figure}[t]
\begin{center}
\includegraphics[width=0.8\textwidth]{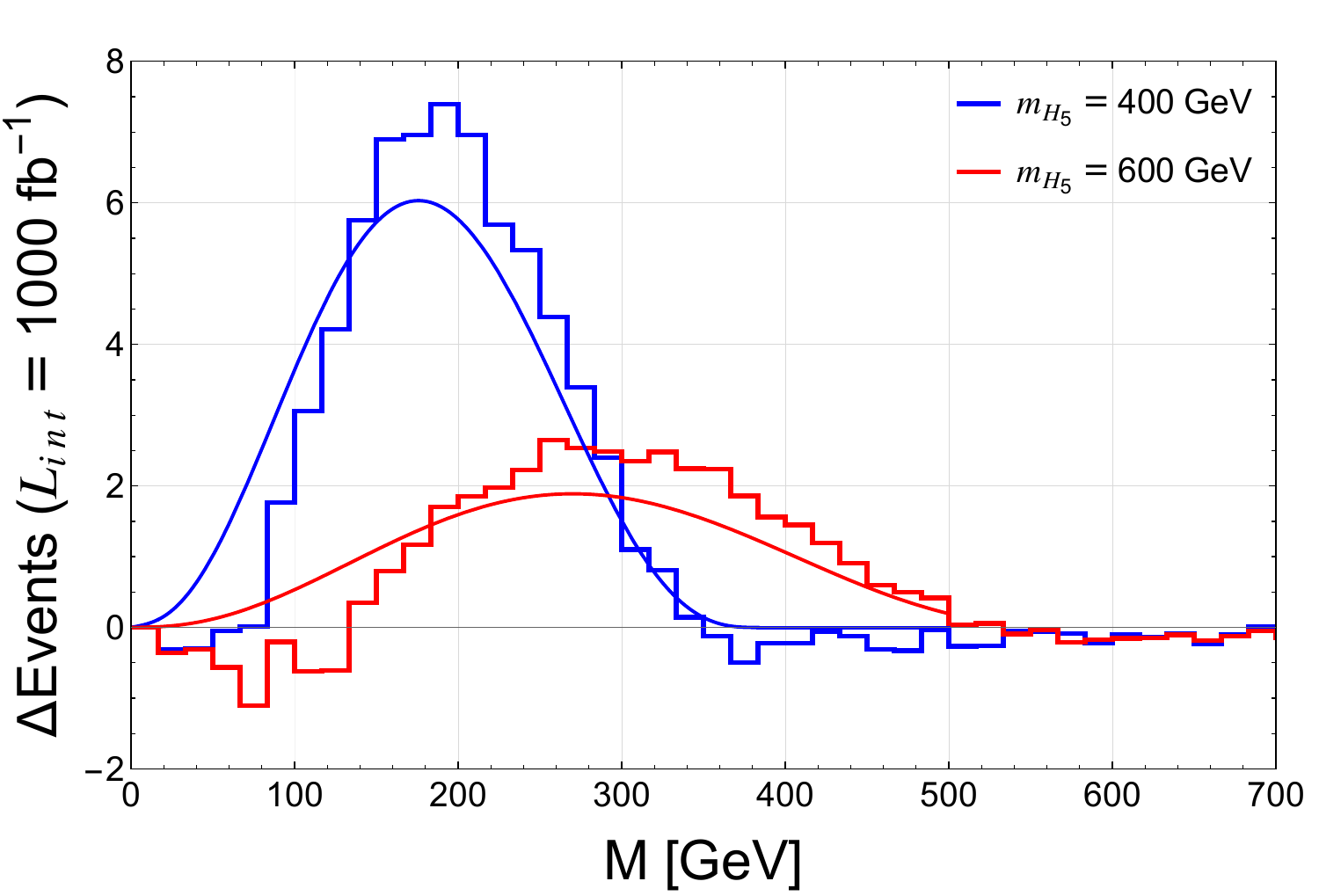}
\caption{Invariant mass distributions in the difference between the signals in the GM model and the SM for $m_{H_5} = 400~\mathrm{GeV}$ (blue) and  $600~\mathrm{GeV}$ (red).  The smooth curves and histograms represent the analytical results and simulations using MG5, respectively.}
\label{fig: Minv_comparison}
\end{center}
\end{figure}

In Fig.~\ref{fig: Minv_comparison}, we show $f_\mathrm{inv}(M)$ and the result of the simulations using MG5 by the smooth curves and the histograms, respectively, in the cases of $m_{H_5} = 400~\mathrm{GeV}$ (blue) and $600~\mathrm{GeV}$ (red). 
We have generated $10^6$ events with $\sqrt{s} = 2~\mathrm{TeV}$ in MG5. 
The total number of events in Fig.~\ref{fig: Minv_comparison} is normalized so that the integrated luminosity is $1000~\mathrm{fb^{-1}}$.
The difference between the curve and the histogram is not only due to statistical fluctuations, but also because $f_\mathrm{inv}(M)$ does not include the effects of kinematical cuts in Eq.~(\ref{eq: kinematica_cut}), the interference terms, and the contribution from $H_5^0$. 
The match between the analytical result and the simulation becomes better if these effects are included.
Here, we neglect them to reduce the computational cost and use $f_\mathrm{inv}(M)$ as the approximate theoretical prediction.

To find the best-fit value of $m_{H_5}$, we minimize the $\chi^2$ for $m_{H_5}$;
\begin{equation}
\chi^2(m_{H_5}^{}) = 
    \sum_{i}
    \biggl(\frac{ y_i -  f_\mathrm{inv}(x_i) \Delta M }{ \sigma_i }\biggr)^2, 
\end{equation}
where $i$ labels the bins of the invariant mass distribution and $\Delta M$ denotes the bin width.
In the $i$-th bin, $x_i$, $y_i$, and $\sigma_i$ are the center value of $M$, the difference between the event numbers in the GM model and the SM, and the error of the event number, respectively. 
We estimate $\sigma_i$ by assuming that the event numbers in the SM and the GM model independently follow the Poisson distribution. 
Then the difference $y_i$ should follow the Skellam distribution~\cite{Skellam:1946}. 
The mean value and the variance in the Skellam distribution are given by $\mu_1-\mu_2$ and $\mu_1+\mu_2$, respectively, where $\mu_1$ and $\mu_2$ are the mean values of each Poisson-distributed variable.  
The total event number $N_\mathrm{ev}$ is evaluated by $(\sigma_\mathrm{GM}-\sigma_\mathrm{SM}) {\cal L}$, where $\sigma_\mathrm{GM}$ and $\sigma_\mathrm{SM}$ are the cross section in the GM model and the SM, respectively, and $\cal L$ is the integrated luminosity. 
We use MG5 to evaluate $\sigma_\mathrm{GM}$ and $\sigma_\mathrm{SM}$ and assume $\mathcal{L} = 1000~\mathrm{fb^{-1}}$.

\begin{figure}[t]
    \centering
    \includegraphics[width=0.9\linewidth]{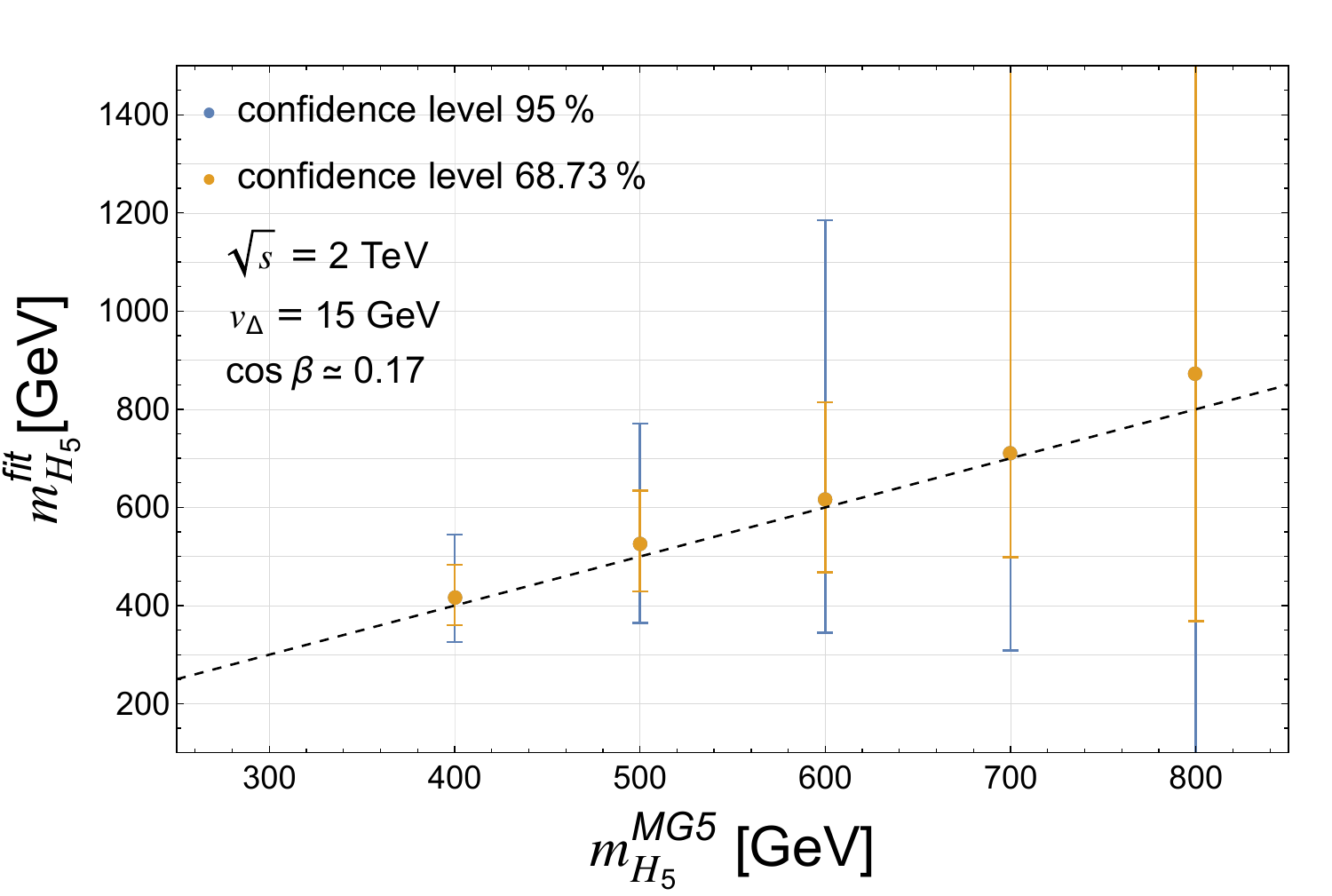}

    \caption{Best-fit $m_{H_5}$ for various masses set in MG5 for the invariant mass distribution of charged leptons of the signal process for $\sqrt{s}=2$~TeV and $v_{\Delta}=15$~GeV ($\cos \beta \simeq 0.17$). The dashed line has a slope of unity.}
    \label{fig: fitting_result}
\end{figure}

Let $m_{H_5}^\mathrm{fit}$ and $m_{H_5}^\mathrm{MG5}$ be the best-fit value in the $\chi^2$ fit and the input value of $m_{H_5}$ in MG5, respectively. 
We show in Fig.~\ref{fig: fitting_result} the correlation between $m_{H_5}^\mathrm{fit}$ and $m_{H_5}^\mathrm{MG5}$ for the input values $m_{H_5}^\mathrm{MG5} = 400$, 500, 600, 700, and 800~GeV. 
The orange and blue bars represent errors at $68.73\%$ CL and $95\%$ CL, respectively. 
The error bars are determined by $|\chi^2(m_{H_5}) - \chi^2_\mathrm{min}| < \Delta \chi^2$, where $\chi_\mathrm{min}^2 = \chi^2(m_{H_5}^\mathrm{fit})$, and $\Delta \chi^2$ is $1.0$ ($3.84$) for $68.73\%$ CL ($95\%$ CL).

We see that the best-fit value is close to $m_{H_5}^{MG5}$, and the latter is included in $1\sigma$ region of the best-fit value at all the points in Fig.~\ref{fig: fitting_result}. 
The error bars are larger for heavier $H_5^{++}$ because the effect of the on-shell $H_5^{++}$ is smaller for the fixed beam energy.  
In the lower mass region, we can expect to obtain the mass information by using this method. 
For heavy $H_5^{++}$ masses, on the other hand, the errors become too large to extract the mass.  
For example, the $2\sigma$ range for $m_{H_5}^\mathrm{MG5}=800~\mathrm{GeV}$ includes $m_{H_5} = 0$.

We have verified that a larger integrated luminosity reduces the errors in the fitting. 
For example, using $\mathcal{L} = 10^4~\mathrm{fb^{-1}}$, the $1~\sigma$ error is reduced to $40~\%$ of that in Fig.~\ref{fig: fitting_result}. 
The mass determination thus becomes more precise for larger integrated luminosities.

Finally, we comment on the mass determination method using the hadronic decay modes: $H_5^{++} \to W^+W^+ \to 4j$.  Although the doubly charged Higgs boson mass can, in principle, be more precisely determined by measuring the invariant mass of the two jets, the electric charge information, however, is less clear in this channel. 
Once $H_5^{++}$ is discovered as a doubly charged particle in the leptonic decay mode with some mass information, as proposed in this work, the hadronic decay modes can be further employed for more precise mass determination.

\section{Same-sign colliders vs. Opposite-sign colliders}
\label{sec: opposite-sign_colliders}

In this section, we compare the single $H_5^{++}$ production at same-sign lepton colliders and pair production at opposite-sign lepton colliders. 
We consider two processes for the pair productions: (i) the production via the single electroweak gauge boson $\gamma^\ast/Z^\ast \to H_5^{++} H_5^{--}$ and (ii) the production via the vector boson fusion $W^+ W^- \to H_5^{++} H_5^{--}$.

\begin{figure}[t]
    \centering
    \includegraphics[width=0.9\linewidth]{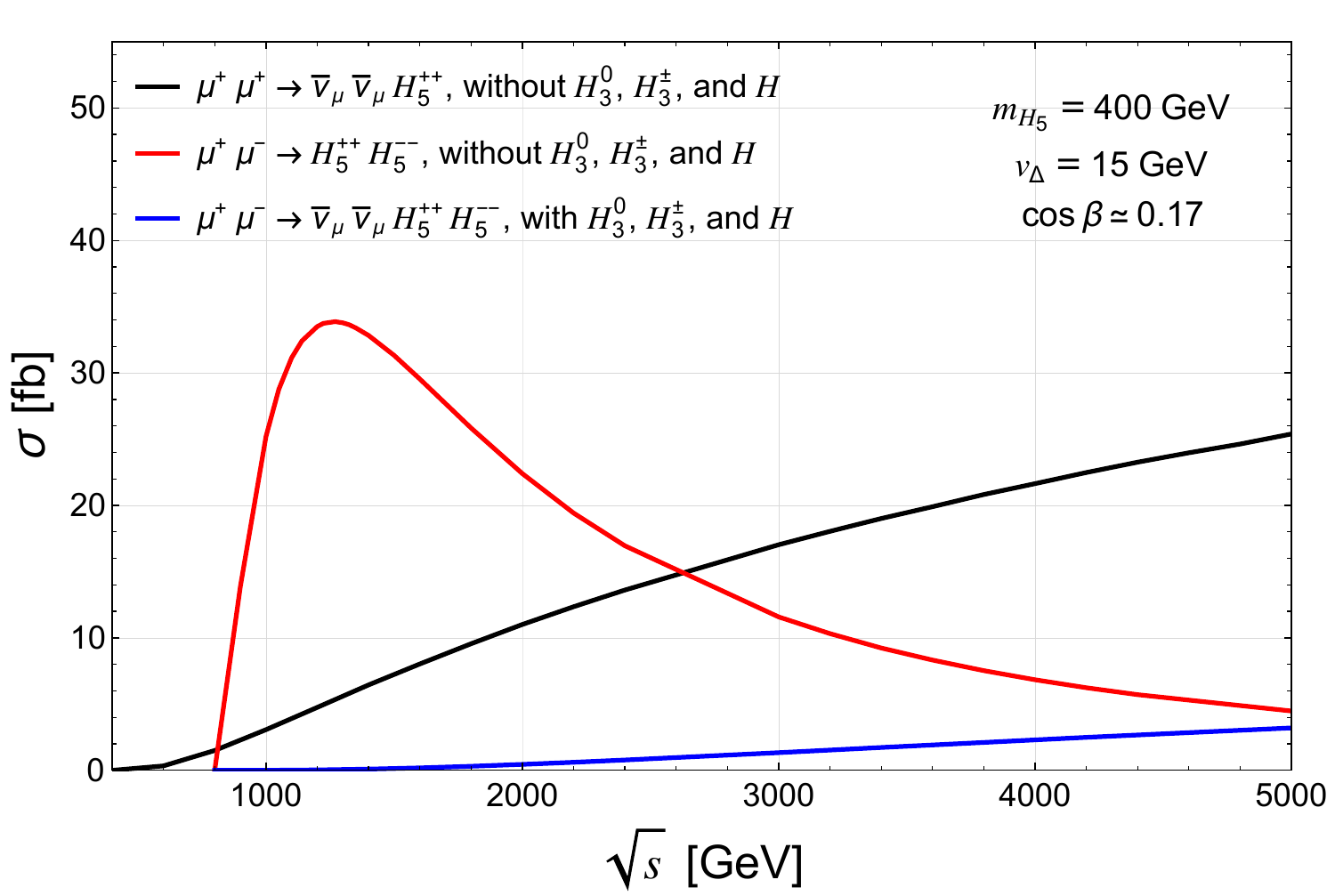}
    \caption{$\sigma$ vs. $\sqrt{s}$ of single and pair productions of doubly charged Higgs bosons with $m_{H_5}=400$~GeV and $v_{\Delta}=15$~GeV ($\cos \beta \simeq 0.17$).} 
    \label{opp}
\end{figure}

In Fig.~\ref{opp}, we show the cross section of the above three production processes as a function of the beam energy, fixing $m_{H_5} = 400~\mathrm{GeV}$ and $v_\Delta = 15~\mathrm{GeV}$. 
We assume $\mu^+ \mu^+$ and $\mu^+ \mu^-$ beams; however, the result is the same in the case of $e^+ e^+$ and $e^+ e^-$ beams, respectively. 
In the single production (the black curve), we neglect the contributions of the triplet and the heavy singlet.  
On the other hand, in the pair production via the $W$ boson fusion (the blue curve), we include the triplet contribution to cancel the quadratic growth at high energies due to the longitudinal $W$ bosons~\cite{Lee:1977eg}.
The mass of the triplet is set as $m_{H_3} \simeq 685~\mathrm{GeV}$. 
The triplet and heavy singlet do not participate in the pair production via $\gamma^\ast/Z^\ast$.

Below $\sqrt{s} < 2m_{H_5}$, only the single production is available. 
At $\sqrt{s} \simeq 2m_{H_5}$, the pair production processes are open.
The cross section of the $H_5^{++} H_5^{--}$ production reaches the peak at $\sqrt{s} \simeq 1.2~\mathrm{TeV}$ (the red curve). 
At higher energies, it decreases due to the suppression by the s-channel propagator. 
The cross section of single production logarithmically increases at high energies. 
At $\sqrt{s} \gtrsim 2.6~\mathrm{TeV}$, the single production gives a larger cross section than the $H_5^{++} H_5^{--}$ production. 
The pair production with a neutrino pair is not effective for $\sqrt{s} \lesssim 5~\mathrm{TeV}$.

Consequently, the single production is more important at higher energies. Although the pair production via $\gamma^\ast/Z^\ast$ is dominant at low energies, the single production has an experimental advantage. 
With the current experimental technology, high-energy $\mu^+$ beams are more feasible than $\mu^-$ beams to reach the high-energy frontier~\cite{Hamada:2022mua}.

\section{Conclusions}
\label{sec: conclusion}

In this paper, we have investigated the single production of the doubly charged Higgs boson in the GM model at the same-sign lepton colliders. 
We have considered the scenario where the triplet VEV $v_\Delta = \mathcal{O}(10)~\mathrm{GeV}$ and thus $H_5^{\pm\pm}$ predominantly decays into a pair of $W$ bosons. 
Hence, the signal is the $W$ boson scattering distorted by the resonant production of $H_5^{\pm\pm}$.

We have focused on the leptonic decays of the produced $W$ bosons. 
In this case, we cannot use the transverse mass of the decay product to obtain the mass of $H_5^{\pm\pm}$ because some neutrinos are radiated from the initial-state charged leptons. 
Instead, we have proposed a method that utilizes the invariant mass of the charged leptons. 
By employing the analytical formula of the invariant mass distribution in the $H_5^{\pm\pm}$ decay, we have performed $\chi^2$ fits and have found the best-fit value of the mass for various input values in the numerical simulations using MG5.
We have demonstrated that this method is effective in extracting the mass information of $H_5^{\pm\pm}$, provided it is not too heavy.

We have also compared the single production at same-sign colliders with the pair production at opposite-sign colliders. 
At high energies, the single production is important because of its logarithmic growth behavior. 
Although the pair production via $\gamma^\ast/Z^\ast$ is dominant at low energies, $\mu^+\mu^+$ colliders have experimental potential and advantages to reach higher energies. 
Therefore, the single $H_5^{++}$ production at $\mu^+ \mu^+$ colliders would play a significant role in future searches for the doubly charged Higgs boson.  Though we have fixed the triplet VEV at 15~GeV in most of our numerical analyses, the results can be readily applied to other values by appropriate scaling.  Moreover, they can also be applied to the HTM, although its triplet VEV can be at most a few GeV.

\section*{Acknowledgments}

We express our sincere thanks to Takeo Moroi and Koichi Hamaguchi for their contributions at the early stage of this project and useful comments on the manuscript.  C.-W.C. would like to thank the hospitality of the High Energy Physics Theory Group at the Department of Physics, University of Tokyo, during his visit in the spring of 2024, when this work was initiated.  This work was performed in part at the Aspen Center for Physics, which is supported by a grant from the Simons Foundation (1161654, Troyer).  This work was supported by the National Science and Technology Council under Grant Nos. NSTC-111-2112-M-002-018-MY3 and NSTC-113-2811-M-002-043.

\appendix

\section{The distribution of the dilepton invariant mass in the decay of $H_5^{++}$}
\label{app: finv}

In this Appendix, we present a detailed derivation of the distribution function $f_\mathrm{inv}(M)$, where $M$ is the invariant mass of the pair of charged leptons in $H_5^{++} \to W^+ W^+ \to \ell^+ \ell^{\prime +} \nu_\ell \nu_{\ell^\prime}$. 
We assume that $H_5^{++}$ is heavier than $2m_W$ so that the production of the $W$ bosons dominates the phase space of the decay process. 
Then, we can use the narrow-width approximation and separate the process into two subprocesses (i) $H_5^{++} \to W^+ W^+$ and (ii) $W^+ \to \ell^+ \nu_\ell (\ell^{\prime +} \nu_{\ell^\prime}^{})$.

The distribution function is given by
\begin{align}
\label{eq: def_finv}
f_\mathrm{inv}(M) = & N_\mathrm{ev} 
    \sum_{\sigma_1, \sigma_2} 
    \int \mathrm{d}^3 q_1 \mathrm{d}^3 q_2 
    F(q_1, \sigma_1; q_2, \sigma_2) 
    \int\mathrm{d}^3 p_\ell \mathrm{d}^3 p_{\ell^\prime}
    f_\ell (p_\ell; q_1, \sigma_1) 
    f_{\ell^\prime} (p_{\ell^\prime}; q_2, \sigma_2)
    \nonumber \\
    & \times 2 M \delta \bigl(M^2 - (p_\ell + p_{\ell^\prime})^2 \bigr), 
\end{align}
where $q_i$ and $\sigma_i$ ($i=1,2$) are the momenta and the polarizations of the $i$-th $W^+$ boson, and $p_{\ell}$ and $p_{\ell^\prime}$ are the momenta of $\ell^+$ and $\ell^{\prime +}$. 
$F(q_1,\sigma_1; q_2, \sigma_2)$ is the distribution function of $q_1$ and $q_2$ in the decay $H_5^{++} \to W^+(\sigma_1) W^+ (\sigma_2)$.  $f_\ell (p_\ell; q, \sigma)$ is the distribution function of the lepton momentum $p_\ell$ in the leptonic decay of the $W^+$ boson with the momentum $q_i$ and the polarization $\sigma_i$. 
These distribution functions are normalized by 
\begin{align}
\label{eq: normalization_F}
\sum_{\sigma_1, \sigma_2} \int \mathrm{d}^3 q_1 \mathrm{d}^3 q_2 \, F(q_1, \sigma_1; q_2, \sigma_2) = 1, 
\quad \int \mathrm{d}^3 p_\ell \, f(p_\ell, q_i, \sigma_i) = 1.
\end{align}
Hence, $f_\mathrm{inv}(M)$ satisfies
\begin{align}
\int \mathrm{d}M f_\mathrm{inv}(M) = N_\mathrm{ev}, 
\end{align}
where $N_\mathrm{ev}$ is the total number of events. 
We note that Eq.~(\ref{eq: def_finv}) is also available in the case of $\ell = \ell^\prime$ because the momentum exchange of the leptons is canceled by the symmetry factor $1/2$ in the phase space integral:
\begin{align}
& \sum_{\sigma_1, \sigma_2} \int \mathrm{d}^3q_1 \mathrm{d}^3q_2 \, F(q_1,\sigma_1;q_2,\sigma_2) 
\nonumber \\
& \times \frac{ 1 }{ 2 } \int \mathrm{d}^3p_{\ell_1} \mathrm{d}^3 p_{\ell_2}
    \Bigl\{ f_\ell (p_{\ell_1};q_1,\sigma_1)
        f_\ell (p_{\ell_2};q_2,\sigma_2) 
    + f_\ell (p_{\ell_2};q_1,\sigma_1)
        f_\ell (p_{\ell_1};q_2,\sigma_2)
    \Bigr\} 
\nonumber \\[5pt]
= & \sum_{\sigma_1, \sigma_2} \int \mathrm{d}^3q_1 \mathrm{d}^3q_2 \, F(q_1,\sigma_1; q_2,\sigma_2) 
\int \mathrm{d}^3p_{\ell_1} \mathrm{d}^3 p_{\ell_2}\, f_\ell (p_{\ell_1};q_1,\sigma_1)
        f_\ell (p_{\ell_2};q_2,\sigma_2),  
\end{align}
where we have used $F(q_1,\sigma_1;q_2,\sigma_2) = F(q_2,\sigma_2;q_1,\sigma_1)$.

As stated in Sec.~\ref{sec: finv}, $f_\mathrm{inv}$ is Lorentz invariant and, thus, can be evaluated in any coordinate system. 
We employ the rest frame of $H_5^{++}$, where the produced $W^+$ bosons propagate along the $z$ axis. The momenta $q_1^\mu$ and $q_2^\mu$ are then given by
\begin{align}
q_1^\mu = (E,0,0,p), \quad q_2^\mu = (E,0,0,-p), \quad p>0.
\end{align}
The momenta of the leptons are given by
\begin{align}
p_j^\mu = (E_j, p_j^\perp \cos \theta_j, p_j^\perp \sin \theta_j, p_{j z}^{}), 
\end{align}
where $j=\ell$ and $\ell^\prime$, and we neglect the lepton mass.

First, we evaluate $F(q_1, \sigma_1; q_2, \sigma_2)$, which is given by
\begin{align}
F(q_1, q_2; \sigma_1, \sigma_2) 
    = & \frac{1}{\Gamma_{H_5} }
    \frac{ 1 }{ 2m_{H_5} } \frac{ 1 }{ (2\pi)^3 2E } \frac{ 1 }{ (2\pi)^3 2 E} 
    \Bigl|\mathcal{M}_{H_5}^{(\sigma_1, \sigma_2)} \Bigr|^2 
    \nonumber \\
    & \times (2\pi)^4 \delta^{(4)} (P - q_1 - q_2), 
\end{align}
where $P = (m_{H_5}^{}, 0, 0, 0)$, $\mathcal{M}_{H_5}^{(\sigma_1, \sigma_2)}$ is the amplitude of the decay process, and $\Gamma_{H_5}$ is the partial decay width of the process with the spin sum:
\begin{align}
\Gamma_{H_5} = \sum_{\sigma_1, \sigma_2}\frac{1}{2m_{H_5}} \int\mathrm{d}^3 q_1 \mathrm{d}^3 q_2 \Bigl| \mathcal{M}_{H_5}^{(\sigma_1, \sigma_2)} \Bigr|^2 (2\pi)^4 \delta^{(4)} (P - q_1 - q_2). 
\end{align}
The prefactor $1/\Gamma_{H_5}$ is for the normalization in Eq.~(\ref{eq: normalization_F}).

A straightforward calculation leads to 
\begin{align}
\begin{split}
F(q_1, \pm; q_2, \mp) 
&= 
\frac{ 8 }{ \pi m_{H_5}^2 }
    \frac{ r^2 }{ \sqrt{1-4r} } \frac{ 1 }{ 1-4r+12r^2}
    \delta^{(4)}(P - q_1 - q_2), 
    \\
F(q_1, L; q_2, L) 
&= 
\frac{ 2 }{ \pi m_{H_5}^2 }
    \frac{ 1 }{ \sqrt{1-4r} } \frac{ (1-2r)^2 }{ 1-4r+12r^2}
    \delta^{(4)}(P - q_1 - q_2),     
\end{split}
\end{align}
where $r = m_W^2 / m_{H_5}^2$, and $\sigma_i = +$, $-$, and $L$ denote the right-handed, left-handed, and longitudinal polarizations, respectively. $F(q_1,\sigma_1;q_2,\sigma_2)$ is zero for other combinations of the polarizations.

Next, we consider the momentum distribution in the leptonic decay of the $W^+$ boson with a fixed polarization;  
\begin{align}
f_\ell (p_\ell; q_1, \sigma_1)  
    = \frac{ 1 }{ \Gamma_{W}^{q_1,\sigma_1} }
    \sum_\mathrm{spin} \frac{ 1 }{ 2 E } \int & 
    \frac{ \mathrm{d}^3 p_\nu }{ (2\pi)^3 2 E_\nu }
    \frac{1}{ (2\pi)^3 2 E_\ell }
    \Bigl| \mathcal{M}_W^{q_1,\sigma_1} \Bigr|^2 
    \nonumber \\
    & \times (2\pi)^4 \delta^{(4)}(q_1 - p_\ell - p_\nu), 
\end{align}
where $p_\nu$ and $E_\nu$ are the momentum and energy of the neutrino, respectively, $\mathcal{M}_W^{q_1,\sigma_1}$ is the amplitude of the process, $\sum_\mathrm{spin}$ represents the spin sum in the final state, and  $\Gamma_W^{q_1,\sigma_1}$ is the partial decay width given by
\begin{align}
\Gamma_{W}^{q_1,\sigma_1} = 
    \sum_\mathrm{spin} \frac{ 1 }{ 2E } \int \frac{\mathrm{d}^3 p_j }{ (2\pi)^3 2 E_\ell }
    \frac{ \mathrm{d}^3 p_\nu }{ (2\pi)^3 2 E_\nu }
    \Bigl| \mathcal{M}_W^{q_1,\sigma_1} \Bigr|^2 
   (2\pi)^4 \delta^{(4)}(q_1 - p_\ell - p_\nu). 
\end{align}

Using the above formulas, we obtain 
\begin{align}
\begin{split}
& f_\ell(p_\ell; q_1, +) = \frac{3}{8\pi p^3 E_\ell} \bigl(E-p-2E_\ell \bigr)^2 
    \delta \biggl(p_{\ell z} - \frac{2EE_\ell - m_W^2}{2p} \biggr), 
    \\ 
& f_\ell(p_\ell; q_1, -) = \frac{3}{8\pi p^3 E_\ell} \bigl(E+p-2E_\ell\bigr)^2 
    \delta \biggl(p_{\ell z} - \frac{2EE_\ell - m_W^2}{2p} \biggr),
    \\
& f_\ell(p_\ell; q_1, L) = \frac{3}{4\pi p^3 E_\ell } \bigl( p^2 - (2E_\ell-E)^2 \bigr) 
    \biggl(p_{\ell z} - \frac{2E E_\ell - m_W^2}{2p} \biggr).    
\end{split}
\end{align}
In the same way, we find $f_{\ell^\prime}(p_{\ell^\prime}; q_2, \sigma_2)$ as
\begin{align}
\begin{split}
& f_{\ell^\prime}(p_{\ell^\prime}; q_2, +) = \frac{3}{8\pi p^3 E_{\ell^\prime}} \bigl(E + p-2E_{\ell^\prime} \bigr)^2 
    \delta \biggl(p_{\ell^\prime z}^{} + \frac{2EE_{\ell^\prime} - m_W^2}{2p} \biggr),
    \\
& f_{\ell^\prime}(p_{\ell^\prime}; q_2, -) = \frac{3}{8\pi p^3 E_{\ell^\prime}} \bigl(E-p-2E_{\ell^\prime} \bigr)^2 
    \delta \biggl(p_{\ell^\prime z}^{} + \frac{2EE_{\ell^\prime} - m_W^2}{2p} \biggr),
    \\
& f_{\ell^\prime}(p_{\ell^\prime}; q_2, L) = \frac{3}{4\pi p^3 E_{\ell^\prime} } \bigl( p^2 - (2E_{\ell^\prime}-E)^2 \bigr) 
    \biggl(p_{\ell^\prime z}^{} + \frac{2E E_{\ell^\prime} - m_W^2}{2p} \biggr).
\end{split}
\end{align}

Combining the above distribution functions, we can find the formula for $f_\mathrm{inv}(M)$.
We briefly explain how to perform the integrations. 
The integrations by $q_1$ and $q_2$ are used to handle $\delta^{(4)}(p - q_1 - q_2)$.  The remaining angular integration leads to just $4\pi$.  This corresponds to the freedom to rotate the system. 
The integral by $p_j^\perp$ ($j=\ell,\ell^\prime$) are replaced by the integral by $E_j$:
\begin{align}
\int \mathrm{d}^3 p_j = \int_0^\infty \mathrm{d}E_j \int_{-E_j}^{E_j} \mathrm{d}p_{jz} \int_{-\pi}^\pi \mathrm{d}\theta_j\, E_j, \quad (j=\ell,\ell^\prime).
\end{align}
The integrals by $p_{j z}$ are evaluated by using the delta functions in $f_{j}$. 
The variable $\theta_\ell$ and $\theta_{\ell^\prime}$ are changed to $\Delta = \theta_\ell - \theta_{\ell^\prime}$ and $\theta_{\ell^\prime}$. 
Then, the integrand is independent of $\theta_{\ell^\prime}$, and we obtain
\begin{align}
\int_{-\pi}^\pi \mathrm{d}\theta_
{\ell} 
\int_{-\pi}^\pi \mathrm{d}\theta_{\ell^\prime} 
= 2 \int_0^{2\pi}\mathrm{d}\Delta
\int_{-\pi}^{\pi-\Delta}\mathrm{d}\theta_{\ell^\prime}
= 2\int_{0}^{2\pi}\mathrm{d}\Delta (2\pi - \Delta). 
\end{align}
The integral by $\Delta$ is evaluated by using $\delta\bigl(M^2 - (p_{\ell} + p_{\ell^\prime})^2\bigr)$. 
We note that there are two solutions of $\Delta$ in $[0,2\pi)$ satisfying $M^2 = (p_\ell + p_{\ell^\prime})^2$.

All the delta functions have been evaluated, and the remaining integration variables are $E_\ell$ and $E_{\ell^\prime}$. 
We define new integration variables $x$ and $y$ via 
\begin{align}
E_\ell = \frac{ E + xp }{ 2 }, \quad 
E_{\ell^\prime} = \frac{ E  + y p }{ 2 }, 
\end{align}
with $x$ and $y$ varying in $[-1,1]$. 
The result is given by 
\begin{align}
\begin{split}
f_\mathrm{inv}(M) 
= & \, N_\mathrm{ev} \frac{ 9M }{ \pi m_{H_5}^2}  \frac{ 1 }{ 1 - 4r + 12 r^2 } 
\int_{-1}^{1} \mathrm{d}x 
\int_{-1}^{1} \mathrm{d} y \, \frac{ 1 }{ \sqrt{ \beta^2 - \alpha^2 } } 
\theta (\beta^2 - \alpha^2 )
\\
& \times \Bigl\{ r^2 (1+x)^2 (1+y)^2 + r^2 (1-x)^2 (1-y)^2 
\\
& \hspace{80pt} + (1-2r)^2 (1-x^2) (1-y^2) \Bigr\},    
\end{split}
\end{align}
where 
\begin{align}
\begin{split}
& \alpha =  (1+xy)(1+v^2) + 2v(x+y) - 2 (1+v)^2 \omega,
\\
& \beta = (1-v^2)\sqrt{1-x^2}\sqrt{1-y^2},
\\
& v = \frac{p}{E} = \sqrt{1-4r},
\\
& \omega = \frac{M^2}{M_\mathrm{max}^2} = \frac{ 4M^2 }{ m_{H_5}^2 (1+v)^2}.    
\end{split}
\end{align}

Note that $\beta^2-\alpha^2$ is a quadratic function of $y$:
\begin{equation}
\beta^2 -\alpha^2 = -Ay^2 + By  + C, 
\end{equation}
where 
\begin{align}
\begin{split}
A &= (1+v^2+2vx)^2 ,
\\
B &= -2\bigl( 2v+x+xv^2 \bigr)\bigl(1+v^2+2vx-2(1+v)^2\omega \bigr),
\\
C &= -(2v+x+xv^2)^2 + 4 (1+v^2+2vx)(1+v)^2 \omega - 4 (1+v)^4\omega^2.    
\end{split}
\end{align}
Here, $A$ is non-negative. 
Thus, the integral is reduced to 
\begin{align}
\int_{-1}^1 \mathrm{d}x  \int_{-1}^1 \mathrm{d}y \, \theta(-Ay^2+By+C)  = 
\int_\mathcal{D}\mathrm{d}x \int_{y_-}^{y_+}\mathrm{d}y, 
\end{align}
where 
\begin{align}
y_{\pm} = \frac{ 1 }{ 2 A }\Bigl(B\pm \sqrt{B^2+4AC} \Bigr).
\end{align}
The domain $\mathcal{D}$ is determined so that both $y_+$ and $y_-$ fall within $[-1,1]$.\footnote{Since $-Ay^2+By+C$ is negative at the boundary $y=\pm1$, either both $y_+$ and $y_-$ fall within $[-1,1]$ or both are outside.}
It is given by the following inequalities
\begin{align}
\begin{cases}
1 + v^2 + 2vx - (1+v)^2 \omega > 0,
\\
-2 (1-v)^2 v x^2 + x \bigl\{ -(1-v)^4 + 2\omega (1+v)^2 (1+v^2) \bigr\}
\\
    \hspace{100pt} + (1-v)^2 (1+v^2) + 4 v (1+v)^2 \omega > 0,
    \\
2 (1+v)^2 v x^2 + x \bigl\{ (1+v)^4 - 2\omega (1+v)^2 (1+v^2) \bigr\}
\\
     \hspace{100pt} + (1+v)^2 (1+v^2) - 4 v (1+v)^2 \omega > 0.     
\end{cases}
\end{align}
With some algebra, $\mathcal{D}$ can be reduced to the simple expression
\begin{align}
\mathcal{D} = 
\begin{cases}
-1 < x < 1, & (0 \leq \omega \leq \omega_c),
\\
x_c < x < 1, &  (\omega_c < \omega \leq 1),    
\end{cases}
\end{align}
where 
\begin{equation}
x_c = \frac{ (1+v)^2 \omega - 1 - v^2 }{ 2 v }, \quad 
\omega_c = \frac{ (1-v)^2 }{ (1+v)^2 }.
\end{equation}

Now, the integrals with respect to $x$ and $y$ can be performed straightforwardly. 
The final result is given by
\begin{align}
f_\mathrm{inv}(M) = N_\mathrm{ev} \frac{ 9(1+v) }{ 4 m_{H_5} }  \frac{ \sqrt{\omega} }{ 3 - 2v^2 + 3v^4 } F(v,\omega), 
\end{align}
where $F(v,\omega)$ is defined as 
\begin{align}
F(v,\omega) 
= 
\begin{cases}
    F_L(v,\omega) & (0<\omega<\omega_c),
    \\
    F_H(v,\omega) & (\omega_c < \omega < 1),    
\end{cases}
\end{align}
and $F_L$ and $F_H$ are given by
\begin{align}
F_L = \frac{ (1+v)^2 }{ 4 v^5 } \biggl[ & 
	v \Bigl\{ -3\omega^2 (1+v)^2 (3 + 5v^2 + 5v^4 + 3v^6) 
    \nonumber \\[5pt]
    &\qquad - (1-v)^2 (9 + 7v^2 + 7v^4 + 9v^6)
	\nonumber \\[5pt]
	&\qquad - 4\omega (9 + 8v^2 + 30 v^4 + 8v^6 + 9v^8) \Bigr\}
    \nonumber \\[5pt]
	&+ \frac{1}{2} \Bigl\{ \omega^2 (1+v)^2(3+2v^2+3v^4)^2 
	\nonumber \\[5pt]
	&\qquad + 4\omega(1+v^2)(9-4v^2+22v^4-4v^6+9v^8)
	\nonumber \\[5pt]
	&\qquad +(9+4v^2+38v^4+4v^6+9v^8)(1-v)^2 \Bigr\} 
	\log\left( \frac{ 1 + v }{ 1 - v } \right) \biggr], 
\end{align}

\begin{align}
F_H = \frac{ (1+v)^2 }{ 16 v^5 } \biggl[ & 
	(\omega-1) \Bigl\{ \omega (1+v)^2 \Bigl(27-18v+44v^2-30v^3+82v^4
        \nonumber \\
        & \hspace{100pt} -30v^5+44v^6-18v^7+27v^8 \Bigr)
		\nonumber \\[5pt]
		& \hspace{40pt} + (1-v)^2 \Bigl(27 + 18 v + 28 v^2 + 14 v^3 + 82 v^4 
            \nonumber \\
            & \hspace{105pt} + 14 v^5 + 28 v^6 + 18 v^7 + 
 27 v^8 \Bigr) \Bigr\} 
	\nonumber \\[5pt]
	& - \log \omega \Bigl\{ \omega^2 (1+v)^2 (3 + 2v^2+3v^4)^2 
        \nonumber \\
        & \hspace{40pt} + (1-v)^2 (9+4v^2+38v^4+4v^6+9v^8)
		\nonumber \\
		& \hspace{40pt} + 4\omega(1+v^2)(9-4v^2+22v^4-4v^6+9v^8) \Bigr\} \biggr].
\end{align}

\bibliographystyle{apsrev4-1}
\bibliography{references}

\end{document}